# Evidence-to-Decision (E2D): Structuring Action-Specific Evidence for Early-Phase Oncology Development

Haitao Pan, PhD

Department of Biostatistics, St. Jude Children's Research Hospital, Memphis, TN, USA

Email: Haitao.Pan@stjude.org

**TECHNICAL COMPANION —** A separate Full Technical Supplement provides the complete statistical specification underlying the quantitative demonstration, including model assumptions and derivations, how current evidence is propagated through the action-specific predictive simulations to future-study outcomes, calibration and threshold selection, simulation scenarios, numerical computation, and reproducibility details. These technical details are intentionally not repeated in this clinical co-design review copy.

# Abstract

**Background.** Early-phase oncology studies are often summarized by asking whether a regimen is promising, positive, or worthy of further development. In practice, however, an encouraging early signal can lead to several very different next questions: should we enroll more patients in the same setting, move to a randomized comparison, study the regimen earlier in therapy, focus on a biomarker-defined subgroup, modify dose or schedule, or first collect more evidence? These are not interchangeable versions of a single go/no-go decision. Each requires a different future study, a different level and type of evidence, and a different balance between potential benefit, uncertainty, and risk. This problem is especially important in rare and pediatric oncology, where patient populations are limited and each development decision carries substantial opportunity cost.

**Framework.** We propose the Evidence-to-Decision (E2D) framework to organize early-phase evidence around these specific development decisions. E2D starts by asking what decision the team is making now and what next actions are genuinely under consideration. For each action, the team specifies the future study, defines what would count as success, identifies what evidence would be needed before taking that step, and compares those requirements with the evidence currently available. When the future study differs from the current setting, the assumptions needed to carry information forward are made explicit. Statistical methods can then quantify how much support the current evidence provides for each contemplated study, how sensitive that support is to key assumptions, and what important evidence remains missing. E2D separates this evidentiary assessment from the final multidisciplinary choice of what to do.

**Quantitative demonstration.** To show how statistics can operate within this clinical framework, we constructed a rare pediatric oncology–motivated example using predictive probabilities. We considered larger single-arm expansion in the same setting, a same-setting randomized Phase II study, movement to an earlier-line or frontline randomized study, and a separate de-escalation example. The first three actions were evaluated only after the current study met prespecified efficacy and safety requirements. Action-specific support rules were calibrated to limit false advance in prespecified insufficient scenarios and false discard in prespecified worthwhile scenarios, and the frozen rules were independently evaluated at higher Monte Carlo precision.

**Results.** All four frozen support rules passed their prespecified higher-precision verification criteria. More importantly, the simulations produced clinically distinct sets of supportable next studies rather than a single continuum of “promise.” Under intermediate efficacy with acceptable toxicity ($p_E, = 0.25\ p_T = 0.10$), 24.9% of simulated trials reached the Development Fork; among those Fork-reaching trials, approximately 35% supported both additional same-setting single-arm expansion and same-setting randomized Phase II while not supporting movement to frontline disease. With stronger efficacy and low toxicity, the set of supportable options broadened: all three advancement actions were supported simultaneously in 65.5% of trials at $p_E = 0.40$, $p_T = 0.10$ and in 88.8% at $p_E = 0.55$, $p_T = 0.10$. In a safety stress scenario with strong efficacy but high toxicity ($p_E, = 0.55\ p_T = 0.35$), approximately 60% of Fork-reaching trials supported only additional same-setting expansion, illustrating that strong activity may still be insufficient for larger comparative or context-changing commitments when safety reassurance is weaker. In the separate de-escalation demonstration, all 2,000 replicates supported de-escalation at assumed efficacy losses of 0 and 0.02, whereas none supported it at losses of 0.05 or 0.10.

**Conclusions.** E2D reframes early-phase evidence from a global judgment of whether a treatment is promising to a more clinically specific question: **which next studies are reasonably supported by the evidence available now, and which are not yet supported?** The quantitative demonstration showed that the same development program can reach evidence states in which the appropriate message is to continue learning in the current setting, proceed to randomized evaluation without yet moving frontline, or recognize that several development paths are simultaneously defensible. The framework can also accommodate a fundamentally different objective such as treatment reduction. Predictive modeling provides one quantitative implementation, but the actions under consideration, the evidence needed for each action, the assumptions connecting current and future settings, and the ultimate choice among supportable options remain clinically defined and multidisciplinary.

# 1. Introduction

Early-phase oncology studies are often summarized using broad labels such as *promising*, *positive*, *go*, or *worthy of further study*. These labels are useful shorthand, but they can hide the actual development question. An encouraging early signal does not necessarily point to one obvious next study. Investigators may consider enrolling additional patients in the same setting, initiating a randomized comparison, moving the regimen into an earlier line of therapy, focusing development on a biomarker-defined subgroup, modifying dose or schedule, reducing treatment burden, or first collecting additional evidence before making a larger commitment. These options differ not only in study design, but also in the patients involved, the comparator, the acceptable balance of benefit and risk, the evidence needed before taking the step, and the consequences of getting the decision wrong.

This problem is especially important in rare and pediatric oncology, where patient populations are limited and evidence often accumulates slowly (Cheung et al., 2024). A decision to pursue one study may affect whether another study is feasible, how scarce patients are allocated, how quickly a regimen can be evaluated, and what questions can realistically be answered next. A poorly matched next study may consume substantial time and patient resources without resolving the clinical uncertainty that matters most. The practical question is therefore often not simply whether a regimen appears active or sufficiently safe. It is **what the evidence available now is sufficient to support next**.

Existing early-phase designs appropriately address important trial-level questions such as activity, toxicity, futility, and dose selection (Simon, 1989; Thall et al., 1995). Predictive probability and assurance-based approaches can also estimate the chance that a prespecified future study will meet a defined success criterion (Spiegelhalter et al., 1986; Lee and Liu, 2008; Temple and Robertson, 2021). The remaining challenge is more clinical than computational: **the next action itself is often treated as if it were already known**. In practice, several different development paths may be under discussion at the same time, and the evidence needed to justify one may not be sufficient to justify another. The evidence needed to enroll more patients in the same disease setting is not necessarily the same as the evidence needed to begin a randomized Phase II study. Moving from relapsed/refractory disease into frontline treatment raises additional questions about what can reasonably be carried forward and what needs to be established again in the new setting.

We propose the **Evidence-to-Decision (E2D)** framework to organize early-phase evidence around these specific clinical-development decisions. E2D begins by defining the decision the team is making now and identifying the next actions genuinely under consideration. For each action, investigators specify the future study, define what would count as successful evidence in that study, identify what information they would want before taking that step, and compare those needs with the evidence currently available. When the future setting differs from the current one, the framework also makes explicit what information is being carried forward, what may change, and what remains uncertain. Statistical methods can then quantify how strongly the current evidence supports each contemplated future study and how dependent that support is on the assumptions being made.

The conceptual contribution of E2D is therefore not a new predictive-probability formula. It is the separation of two questions that are often blurred together: **What next studies does the current evidence reasonably support?** and **Which supported option should the development team ultimately choose?** Different actions may require different evidence, rely on different assumptions, and define success differently. The same current evidence may support one next study but not another; under stronger evidence, several development paths may become supportable at the same time. E2D does not force those options into a single ranking or select a statistical winner. The final choice remains a multidisciplinary clinical-development decision.

In this paper, we first describe the E2D clinical decision architecture and an action-specific evidence map for early-phase oncology development (Figure 1). We then show how predictive modeling can provide a quantitative layer within that architecture using a rare pediatric oncology–motivated demonstration. The example considers additional same-setting single-arm experience, a randomized Phase II study in the same setting, movement to an earlier-line or frontline randomized study, and a separate de-escalation question. The purpose of these examples is not to establish universal development rules, but to show how the same early evidence can lead to different sets of supportable next studies once the clinical question for each action has been specified. We then examine what limits support for different actions, what additional evidence might change that support, and how E2D could evolve into a clinically co-designed evidence-planning framework.

## 2. The Clinical Development Decision Problem

At a clinical-development decision point, the practical question is rarely just whether a regimen looks encouraging. The more immediate question is **what the team is considering doing next**. Two teams looking at exactly the same response and toxicity data may therefore be making very different decisions: one may be asking whether to enroll additional patients in the same setting, while another may be asking whether the evidence is mature enough to justify a randomized study or a move into frontline therapy.

An encouraging early-phase signal may reasonably justify gaining more experience in the same disease setting. The same signal may also prompt discussion of a randomized Phase II study. Moving the regimen into earlier-line or frontline therapy, however, is a different question. The patients may differ, expected outcomes on the control treatment may differ, the relevant endpoints may change, and the amount of toxicity that is acceptable may be very different from that in relapsed/refractory disease. Biomarker-enriched development, dose or schedule modification, and treatment de-escalation likewise raise their own clinical questions and require different information before the team is ready to proceed.

These options should therefore not be treated as progressively stronger versions of one generic decision to "continue development." Each represents a different future study and asks a different scientific question. Evidence that is sufficient to justify one next step may still be insufficient for another, even though both decisions arise from the same early-phase dataset. In E2D, the future study being contemplated—not the treatment in the abstract—is what the current evidence must support.

For that reason, the clinical question comes before the statistical calculation. For each plausible next step, the team first asks: What study are we actually considering? What would we need to know before we felt ready to undertake it? Which of that evidence do we already have? What are we carrying forward from the current setting, and what remains uncertain? Only after those questions have been specified should quantitative support be calculated. The candidate actions should come from real clinical-development alternatives, not from the statistical model, and the definition of success should reflect what investigators would genuinely consider meaningful in the contemplated future study.

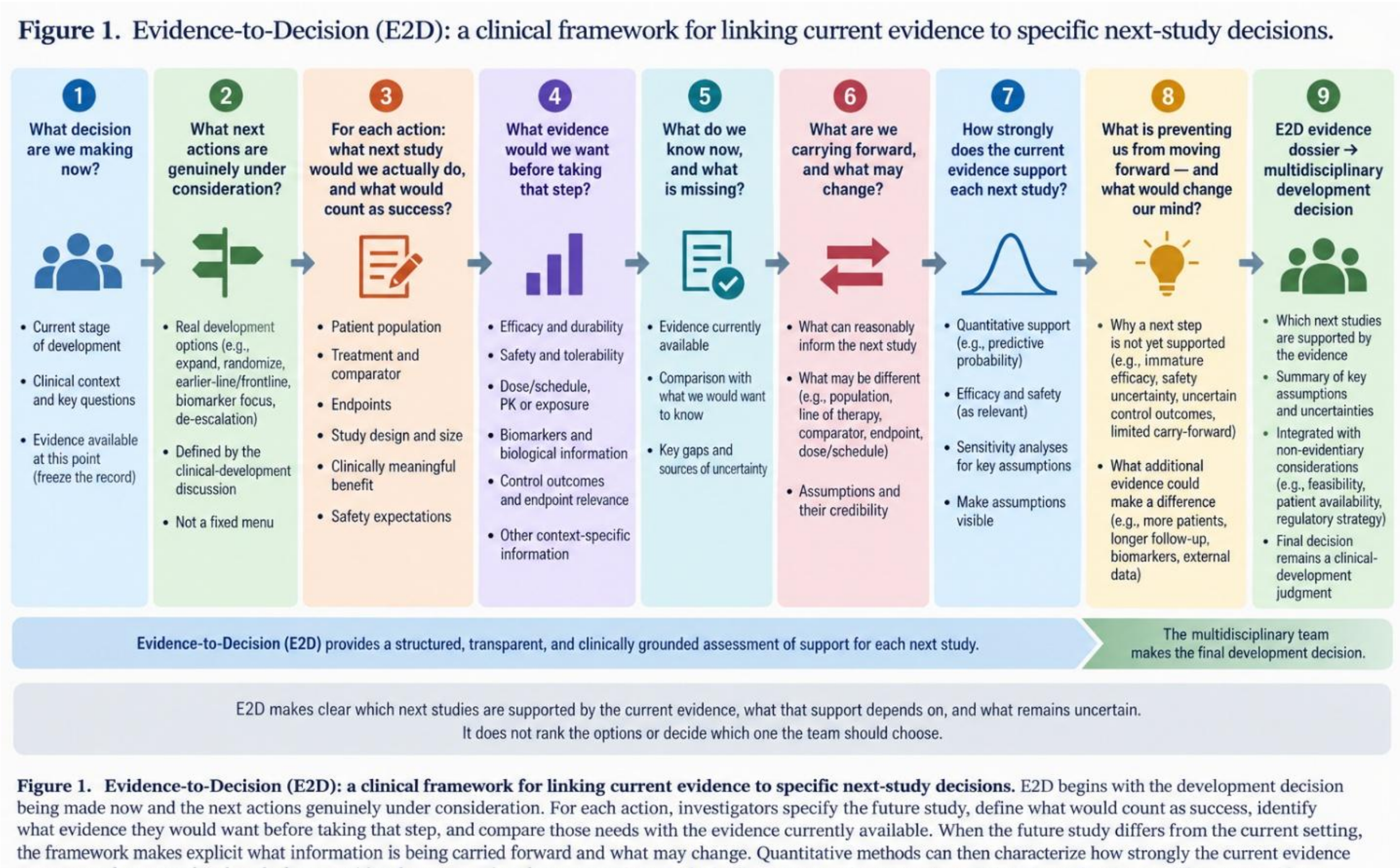

**Figure 1.** Evidence-to-Decision (E2D): a clinical framework for linking current evidence to specific next-study decisions.

**Figure 1. Evidence-to-Decision (E2D): a clinical framework for linking current evidence to specific next-study decisions.** E2D begins with the development decision being made now and the next actions genuinely under consideration. For each action, investigators specify the future study, define what would count as success, identify what evidence they would want before taking that step, and compare those needs with the evidence currently available. When the future study differs from the current setting, the framework makes explicit what information is being carried forward and what may change. Quantitative methods can then characterize how strongly the current evidence supports each contemplated study, how sensitive that support is to key assumptions, and what important evidence remains missing. Several actions may be supportable at the same time. E2D informs multidisciplinary deliberation; it does not rank the actions or determine which one should be chosen.

## 3. The Evidence-to-Decision Framework

### 3.1 Define the clinical-development decision point

E2D begins with a simple question: **what decision is the development team actually facing now?** That decision point may arise after dose selection, completion of an expansion cohort, emergence of an efficacy signal, accumulation of sufficient safety follow-up, or another clinically meaningful stage of development. It does not have to coincide with the formal end of Phase I or with a particular trial label. What matters is that the team is deciding what should reasonably be considered next.

The evidence available at that moment should be recorded explicitly. Depending on the setting, this may include treatment activity, safety and tolerability, response durability, dose and schedule, PK or exposure, pharmacodynamic or biological information, biomarkers, external benchmarks, disease-context information, and data that remain missing or immature. The purpose is to capture **what the team actually knew when the decision was being made**, rather than allowing information learned later to reshape the original decision retrospectively.

### 3.2 Identify candidate actions genuinely under consideration

The next step is to identify the development options the team is genuinely considering at that decision point. E2D does not assume a fixed menu. Depending on the program, those options might include enrolling additional patients in the same setting, beginning a randomized comparison, moving the regimen into an earlier line of therapy, focusing on a biomarker-defined population, modifying dose or schedule, reducing treatment burden, or deliberately collecting additional evidence before making a larger commitment.

The important point is that these options should come from the **clinical-development discussion**, not from the statistical output. The action set should therefore be allowed to change during clinical co-design. An option may turn out to be unrealistic, an important alternative may be missing, or two options that initially appear similar may actually require very different future studies. E2D should reflect those distinctions rather than forcing real development decisions into a predetermined menu.

## 3.3 Specify the future study and future-success event for each action

Once a candidate action has been identified, it should be translated into a concrete future study. The team should specify, as appropriate, who would be studied, what treatment and comparator would be used, what outcomes would matter, how those outcomes would be analyzed, how large the study would be, what magnitude of benefit would be clinically meaningful, and what level of safety reassurance would be required.

This step matters because a development action cannot be evaluated meaningfully in the abstract. “Do a randomized trial” is not yet a sufficiently defined decision. A randomized study in relapsed/refractory disease and a randomized frontline study may involve different patients, control expectations, endpoints, treatment-effect targets, safety expectations, and practical consequences. Likewise, “expand the cohort” or “de-escalate treatment” becomes meaningful only after the future study and the evidence that would count as success have been specified.

For example, additional experience in the same setting might be considered successful if the combined evidence continues to show adequate activity with acceptable toxicity. A randomized study may require a clinically meaningful advantage over the control treatment together with sufficient safety reassurance. A de-escalation study may require preservation of efficacy within an acceptable margin while achieving a meaningful reduction in treatment burden. These definitions differ because the studies are answering different clinical questions.

## 3.4 Define what evidence would be needed before taking each next step

Specifying the future study is only part of the decision. The team must also ask: what evidence would we want to see before we felt comfortable taking that step? The answer will differ by action. Relevant information may include treatment activity and durability, toxicity and tolerability, dose and schedule, PK or exposure, pharmacodynamic or biological information, biomarkers, expected outcomes with the comparator, endpoint relevance, and evidence about how similar or different the future patient population may be.

Not every type of evidence matters equally for every decision. Additional experience in the same population may require confidence that the new patients can reasonably be interpreted together with the current cohort and that assessment methods have not materially changed. A randomized study in the same setting additionally requires a clinically appropriate comparator, a meaningful treatment-effect target, and enough safety experience to justify comparative evaluation. Moving from relapsed/refractory disease into frontline treatment raises a different set of questions: which findings from the current setting are likely to remain informative, which may change in the new population and treatment context, and which need to be established again before proceeding.

Operational considerations also matter, but they serve a different role. Patient availability, competing trials, drug supply, feasibility of randomization, and regulatory strategy may strongly influence whether a supported study can actually be pursued. E2D records these considerations alongside the evidence assessment rather than forcing them into the statistical support calculation.

**Table 1. Illustrative evidence map for common early-phase oncology development decisions**

| Candidate action | Clinical-development question | Contemplated future study | Evidence we would want to review | What else would need to be known or justified? |
|---|---|---|---|---|
| Larger single-arm expansion | Do we need additional experience in the same setting before making a larger commitment? | Additional cohort in the same population, disease setting, dose/schedule, and endpoint context. | Activity; toxicity; response durability; dose feasibility; potentially PK/exposure when relevant. | Are the additional patients sufficiently similar to the current cohort? Have treatment context or assessment methods changed over time? Can the current and new data reasonably be interpreted together? |
| Same-setting randomized Phase II | Is the current evidence sufficient to justify comparative evaluation in the same clinical setting? | Standalone randomized comparison in the same disease setting. | Strength and durability of activity signal; toxicity; dose/schedule; clinically meaningful treatment effect; feasibility of randomization. | What should we expect from the control arm? Is the comparator still clinically appropriate? Is the safety experience mature enough to justify randomization? What magnitude of benefit would be clinically meaningful? |
| Earlier-line / frontline evaluation | Is the evidence sufficient to move the regimen into a different disease setting? | Comparative study in earlier-line or frontline disease. | Current efficacy; safety; biological rationale; dose/exposure information; durability of effect. | Which findings from the relapsed/refractory setting are likely to remain informative in frontline disease? How different are the patients, expected control outcomes, relevant endpoints, and acceptable toxicity in the new setting? |
| Biomarker-enriched development | Should further development focus on a biomarker-defined subgroup? | Enriched expansion or randomized study. | Biomarker definition; assay performance; subgroup activity; biological plausibility; evidence of treatment-effect heterogeneity. | Is the biomarker clearly defined and reliably measured? Is the subgroup common enough for a feasible study? Is there enough evidence that the treatment effect is truly concentrated in this group? |
| Dose / schedule modification | Should the regimen be studied with a different dose, schedule, or exposure strategy? | Dose/schedule optimization or expansion study. | Toxicity; PK/PD; exposure-response; efficacy; cumulative toxicity; feasibility. | Do changes in dose or schedule meaningfully change exposure, activity, or toxicity? Are delayed or cumulative toxicities sufficiently understood? Do we have enough dose/exposure-response information to justify a new strategy? |
| De-escalation / burden reduction | Can treatment burden be reduced while preserving acceptable efficacy? | Noninferiority or treatment-reduction study. | Baseline efficacy; durability; safety; treatment burden; clinically acceptable efficacy loss. | How much loss of efficacy, if any, would be clinically acceptable? How should treatment burden be measured? Is standard-treatment efficacy sufficiently well established to support a de-escalation study? |
| Additional evidence before commitment | What information is preventing an important action from being supportable now? | Additional cohort, longer follow-up, biomarker study, external evidence, or another targeted evidence-generating action. | Depends on the uncertainty limiting the action. | What uncertainty is actually preventing the team from moving forward? Can that uncertainty be reduced within a clinically useful timeframe? Would the additional information realistically change the development decision? |

## 3.5 Compare what we would want to know with what we know now

Once the evidence needed for a particular next step has been identified, E2D compares those needs with the information actually available at the decision point. This distinction is important because a regimen may look encouraging overall while the team still lacks the specific information needed for a particular development move.

For example, the current activity and safety experience may be sufficient to justify enrolling additional patients in the same setting, while still leaving too much uncertainty to justify moving directly into frontline disease. In that case, the important result is not simply that the frontline action is "unsupported." The more useful result is **what is missing**—for example, insufficient durability, immature safety follow-up, uncertainty about dose or exposure, uncertainty about expected frontline control outcomes, or limited confidence that the current treatment effect will remain relevant in the new setting.

E2D therefore treats the evidence gap as part of the output. A useful assessment should tell the team not only which next studies are currently supported, but also what information is preventing another important option from being supported.

## 3.6 Make explicit what we are carrying forward—and what may change

Whenever current evidence is used to inform a future study, some degree of carry-forward is unavoidable. When the future study closely resembles the current one, the assumptions may be relatively modest. When the population, line of therapy, comparator, endpoint, dose or schedule, or clinical context changes, those assumptions become more important.

E2D makes these assumptions visible rather than allowing them to remain hidden inside a statistical model. In practice, some may be supported by existing clinical data, external evidence, or biological rationale. Others may be uncertain and should be examined through sensitivity analyses. Still others may be too weak to justify carrying the current evidence into the new setting at all.

The practical clinical question is therefore not simply whether the current evidence is "transportable." It is: **what from the current setting are we comfortable using to inform the next study, what might reasonably change, and what should be established separately?** Quantitative models can formalize these assumptions and examine their consequences, but the credibility of the assumptions themselves must be judged clinically.

## 3.7 Quantify how strongly the current evidence supports each next study

Once the clinical question, future study, evidence needs, and key assumptions have been specified, quantitative methods can be used to ask: how strongly does the evidence we have now support taking this particular next step?

Predictive probability is one way to answer that question, but it is not the definition of E2D. Other statistical approaches could be used within the same framework as long as they address the clinically specified future study, preserve the relevant uncertainty, and make the assumptions underlying the calculation visible.

The resulting quantitative support should not be interpreted in isolation. For some actions, a single summary probability may be adequate. For others, it may be more useful to see efficacy and safety separately so that the team can understand what is driving or limiting support. When conclusions depend strongly on assumptions about the future comparator, changes in the patient population, or what is being carried forward from the current setting, those dependencies should be shown explicitly through sensitivity analyses.

## 3.8 Identify what is missing and ask "what would change our mind?"

E2D is intended not only to tell the team whether an important next step is currently supported, but also to explain **why it is not yet supported when the answer is no**. The limiting issue may be immature efficacy follow-up, insufficient safety experience, uncertainty about future control outcomes, limited confidence in applying the current evidence to a different disease setting, inadequate biomarker information, or another gap specific to the proposed next study.

These distinctions matter because they imply different responses. If the main uncertainty is response durability, the team may need longer follow-up rather than more patients. If safety experience is immature, additional exposure may be more informative. If expected control performance is uncertain, better external information may be needed. If the concern is whether the current results will remain relevant in frontline disease, additional clinical or biological evidence may be more useful than simply enlarging the current cohort.

This leads naturally to a second question: **what additional information would actually change our willingness to take the next step?** Depending on the decision, the answer might be additional patients, longer follow-up, more mature safety information, PK or exposure data, biomarker characterization, a more credible external benchmark, or another targeted source of evidence. The present framework does not formally choose among these evidence-generating strategies. Its role is first to make clear **which uncertainty is preventing the team from moving forward** and whether that uncertainty can realistically be reduced.

In this sense, "collect more evidence first" is not a failure to make a decision. It can itself be a legitimate development action when the missing information is identifiable, obtainable within a clinically useful timeframe, and likely to change what the team is prepared to do next.

## 3.9 Separate evidentiary support from the final development decision

E2D stops after asking which next studies are reasonably supported by the available evidence. It does **not** determine which supported option the team should ultimately choose.

That distinction matters because the final decision may depend on considerations that are critical but are not the same as evidence about whether a future study is scientifically supported. Patient availability, competing trials, operational feasibility, drug supply, regulatory strategy, clinical urgency, sponsor considerations, investigator judgment, and patient or family priorities may all influence which path is ultimately pursued.

E2D therefore separates two questions:

**Which next studies are supported by the evidence we have now?**

and

**Among the supported options, which one should we actually pursue?**

The first is the primary evidentiary question addressed by E2D. The second remains a multidisciplinary clinical-development judgment. Keeping these questions separate allows statistics to make the evidentiary basis of the decision clearer without transferring decision authority to the model.

Where useful, the non-statistical considerations that influenced the final choice should still be documented alongside the quantitative assessment. Their role is not hidden or diminished; they are simply kept distinct from the calculation of evidentiary support.

# 4. Quantitative Support Within E2D

Once the development team has defined the decision point, the next studies under consideration, what would count as success for each study, what evidence is needed, and what assumptions are required, the quantitative question becomes more focused:

**Given what we know now, how likely is this particular future study to produce the evidence we would regard as successful?**

Predictive probability provides one way to answer that question. The complete statistical specification, calibration procedures, simulation details, and supplementary operating-characteristic results are provided in the separate Technical Supplement.

For action $a$, let $S_a(D_{\text{future}})$denote the prespecified success event for the future study associated with that action. Let $D$denote the evidence available at the current decision point, and let $M_a$denote the assumptions used to connect the current evidence to the contemplated future study. We define the action-specific predictive probability as

$$PP_a(D) = \Pr\{S_a, (D_{\text{future}}) = 1 \mid DM_a\}.$$

This quantity should be interpreted narrowly. It is **not** the probability that a treatment is generally “promising,” nor is it the probability that the clinical team should choose action $a$. It is the probability that the specified future study will satisfy its prespecified success criteria, conditional on the evidence available now and the assumptions used for that particular development action.

This distinction is important because predictive support can change even when the observed current data do not. A change in the future population, comparator, sample size, clinically meaningful effect, safety requirement, or assumptions about what carries forward into the new setting can change the probability of future success. The quantitative result therefore belongs to a **specific contemplated study**, not to the treatment in the abstract.

In the present implementation, the predictive calculation follows five steps. First, the current data update uncertainty about treatment efficacy and unacceptable toxicity. Second, that uncertainty is carried into the contemplated future study using the assumptions specified for that action, including uncertain control expectations where relevant. Third, possible future trial outcomes are generated. Fourth, each simulated future study is evaluated against the success criteria that were defined for that action. Finally, the proportion of simulated studies meeting those criteria estimates the predictive probability.

The candidate actions are evaluated separately and in parallel rather than being placed into a statistical ranking. For larger same-setting single-arm expansion and de-escalation, a single predictive probability summarizes complete future-study success. For the randomized actions, efficacy and experimental-arm safety are retained as separate predictive components, $PP_E$ and $PP_T$. Support for a randomized action requires both prespecified component criteria to be met. A joint probability of complete future success may be reported descriptively, but it is not used as the support gate.

Predictive probability is therefore one quantitative implementation of E2D, not the framework itself. The clinical architecture remains the same if a future application uses different endpoints, time-to-event outcomes, external or hybrid controls, sequential evidence, alternative Bayesian or frequentist models, or another quantitative engine. What remains fixed is the sequence of questions: **what action are we considering, what future evidence would count as success, what information and assumptions connect the present to that study, and how strongly does the current evidence support taking that step?**

# 5. Pediatric Oncology–Motivated Worked Example

To illustrate how E2D can be used in practice, we constructed a rare pediatric oncology–motivated example centered on a familiar development situation: an early-phase study has generated an initial signal of activity and safety, and the team must decide what kind of evidence-generating step, if any, is justified next.

The example is intentionally generic. It does not represent a specific NANT trial and should not be interpreted as prescribing a standard development pathway for pediatric oncology. Its purpose is to create a controlled setting in which the **same current trial evidence** can be examined against several different next-study questions.

After the current study meets its own efficacy and safety requirements, three advancement options are considered in parallel: additional single-arm experience in the same disease setting, a randomized Phase II study in that same setting, and a randomized study in an earlier-line or frontline population. A separate de-escalation example asks a qualitatively different question: whether the evidence is sufficient to justify studying less treatment rather than intensifying development.

These examples were chosen because they represent different types of development decisions. One asks whether more experience is needed before making a larger commitment. Another asks whether the evidence is mature enough for comparative evaluation. A third asks whether findings from the current setting are sufficiently informative to justify movement into a different clinical setting. The de-escalation example asks whether treatment burden might be reduced while preserving acceptable efficacy. The examples are illustrative rather than exhaustive; a real E2D application should include only the actions that the clinical-development team is genuinely considering.

## 5.1 Current trial and the Development Fork

The worked example begins with a simple early-phase trial that serves only as the **current evidence source** for the downstream E2D decisions. It is not intended to define the E2D framework itself. The trial generates binary outcomes for favorable efficacy and unacceptable toxicity, allowing the subsequent development questions to be illustrated without adding unnecessary complexity from the current-study design.

The efficacy component uses a Simon optimal two-stage design (Simon, 1989). The null response probability is 0.20, the alternative is 0.40, the one-sided type I error is at most 0.10, and power is at least 0.80. Twelve patients are enrolled in Stage 1. The study stops for efficacy futility if two or fewer patients respond; otherwise, provided the safety criterion is also satisfied, enrollment continues to a total of 25 patients. Final efficacy passage requires at least eight responses among 25 patients.

Safety is monitored separately using a Bayesian rule with a maximum acceptable unacceptable-toxicity probability of 0.30. At the interim review, the study stops for toxicity when the posterior probability that the toxicity rate exceeds 0.30 reaches the prespecified threshold, corresponding operationally to five or more unacceptable-toxicity events among 12 patients. At final review, safety passes when the posterior probability that toxicity is below 0.30 reaches the prespecified criterion, corresponding to no more than six unacceptable-toxicity events among 25 patients. The detailed calibration of these safety rules is provided in the Technical Supplement.

Enrollment pauses after Stage 1 until efficacy and toxicity assessments are complete. The **Development Fork** is reached only if the study continues after the interim review and ultimately meets both the final efficacy and safety requirements. Only then are the three post-trial advancement options evaluated. Trials that do not reach the Fork receive no support for those three actions. The de-escalation example is separate and does not use this gate.

The Development Fork should be viewed as an **illustrative decision point**, not as a required feature of E2D. In a real application, E2D could begin after a different clinical milestone—for example, after dose selection, after a cohort expansion, after longer safety follow-up, or at another point when the development team is deciding what to do next.

5.2 Larger same-setting expansion: learning more before a larger commitment

The first action asks a relatively modest development question: is the current evidence strong enough to justify gaining additional experience in the same population and disease setting before making a larger commitment? Compared with the randomized or frontline options, this action changes relatively little. The patients, treatment, dose and schedule, and endpoint context remain the same, so the main clinical concern is whether the additional cohort can reasonably be interpreted together with the patients already treated.

In the illustrative implementation, 20 additional patients are enrolled using the same eligibility context, disease setting, dose, schedule, and efficacy and toxicity endpoints as the current study. The current efficacy and toxicity distributions are carried forward without modeled between-cohort change, and the current and additional patients are analyzed together. In a real application, this assumption would not be automatic: the team would need to consider whether patient mix, assessment methods, treatment delivery, supportive care, or other aspects of the clinical context had changed enough to make pooling questionable.

Successful combined evidence requires both adequate activity and acceptable toxicity:

$$\mathbf{Pr}(\boldsymbol{p_E} > \mathbf{0.20} \mid \boldsymbol{D}_{\text{combined}}) \geq \mathbf{0.90}$$

and

$$\mathbf{Pr}(\boldsymbol{p_T} < \mathbf{0.30} \mid \boldsymbol{D}_{\text{combined}}) \geq \mathbf{0.90}.$$

The predictive probability $\boldsymbol{PP}_{\mathbf{LSA}}$ estimates the chance that both conditions will be met if the additional cohort is undertaken. After the Development Fork is reached, this action is considered supported when

$$\boldsymbol{PP}_{\mathbf{LSA}} \geq \mathbf{0.20}.$$

The 0.90 posterior criteria define what would count as successful combined evidence after the additional patients are enrolled; the calibrated 0.20 predictive threshold addresses a different question—whether the evidence available now is sufficient to support undertaking that additional expansion.

The clinical point is that “enroll more patients in the same setting” is a legitimate development decision in its own right. It should not be treated merely as a weaker version of the question of whether the regimen is ready for randomization. In some evidence states, learning more in the current setting may be supportable even when a larger comparative commitment is not.

## 5.3 Same-setting randomized Phase II: is the evidence mature enough for comparative evaluation?

The second action asks a different question: **has the current evidence matured enough to justify moving from single-arm activity evidence to a randomized comparison in the same clinical setting?**

This step requires more than evidence that the regimen appears active. The experimental treatment must have sufficient efficacy and safety support, the comparator must remain clinically relevant, expected control performance must be credible, and the treatment difference that would justify a positive randomized study must be specified in advance. These requirements make randomization a different evidence problem rather than simply the next rung on a fixed development ladder.

The illustrative future study is a standalone randomized Phase II trial of 60 patients, with 30 patients per arm. Patients from the current early-phase study inform prediction but are not included in the formal analysis of the future randomized trial. For prediction, experimental efficacy and toxicity are assumed not to change between the current and future same-setting cohorts. Uncertainty about future control efficacy is represented by a Beta(2,8) distribution, with mean 0.20 and effective sample size 10. This benchmark is used to represent uncertainty about future control performance; it is not used as an informative prior in the formal analysis of the future trial.

The future trial itself is analyzed using neutral Beta(1,1) priors. Successful randomized evidence requires both a clinically meaningful efficacy difference and acceptable experimental-arm toxicity:

$$\Pr(p_E - p_C > 0.15 \mid D_{\text{future}}) \geq 0.90$$

and

$$\Pr\left(p_{T,E} < 0.30 \mid D_{\text{future},T,E}\right) \geq 0.90.$$

The first criterion asks whether the future data support an experimental-control response difference greater than 0.15; the second asks whether experimental-arm toxicity remains acceptably low. Efficacy and safety are retained as separate predictive components so that one cannot compensate for failure of the other. After Development Fork passage, support for this action requires

$$PP_E \geq 0.16$$

and

$$PP_T \geq 0.41.$$

A joint future-success probability may be reported descriptively, but it is not used as the decision gate.

The numerical thresholds are demonstration-specific. The broader clinical lesson is that moving from single-arm evidence to randomization introduces new questions that were not required for simple same-setting expansion: **What should we expect from the control arm? What magnitude of benefit would matter clinically? Is the safety experience mature enough for comparative evaluation?** E2D makes those additional requirements explicit before asking whether the randomized study is supported.

## 5.4 Earlier-line or frontline randomized Phase II: what can we reasonably carry forward?

The third action considers moving a regimen from relapsed/refractory disease into an earlier-line or frontline setting. Clinically, this is not simply a stronger version of the same development decision. The patients may differ, baseline prognosis may change, the comparator may be different, endpoints may have different meaning, and the acceptable balance between potential benefit and toxicity may shift substantially. The key question therefore becomes: **which findings from the relapsed/refractory setting can reasonably inform the frontline study, and which need to be established again?**

The quantitative demonstration makes this carry-forward assumption explicit rather than leaving it implicit. Current relapsed/refractory efficacy and toxicity uncertainty is combined with uncertain control benchmarks for both the relapsed/refractory and frontline settings. The reference control distributions are Beta(2,8), with mean 0.20, for relapsed/refractory disease and Beta(5,5), with mean 0.50, for frontline disease; both have effective sample size 10.

For efficacy, the reference model assumes that the relapsed/refractory log-odds treatment effect is preserved when moving to frontline disease:

$$\text{logit}(p_{E,F}) = \text{logit}(p_{C,F}) + \lambda[\text{logit}(p_{E,R}) - \text{logit}(p_{C,R})],$$

with

$$\lambda = 1.$$

Safety is carried forward separately using

$$\text{logit}(p_{T,F}) = \text{logit}(p_{T,R}) + \eta_T,$$

with

$$\eta_T = 0.$$

These values are **illustrative development assumptions**, not empirical claims that efficacy and toxicity truly behave this way when neuroblastoma treatment moves from relapsed/refractory to frontline disease. Their purpose is to show how an assumption about what carries forward can be made explicit and its consequences evaluated quantitatively. A real application would require clinical judgment, relevant empirical evidence where available, and sensitivity analyses around assumptions that remain uncertain.

The contemplated future study is a standalone randomized trial of 60 patients, with 30 patients per arm. Its formal analysis again uses neutral Beta(1,1) priors. Successful future evidence requires both a clinically meaningful frontline efficacy difference and acceptable experimental-arm toxicity:

$$\Pr(p_{E,F} - p_{C,F} > 0.15 \mid D_{\text{future}}) \geq 0.90$$

and

$$\Pr(p_{T,F} < 0.30 \mid D_{\text{future},T,F}) \geq 0.90.$$

After Development Fork passage, support for the frontline action requires

$$PP_E \geq 0.27$$

and

$$PP_T \geq 0.39.$$

The purpose of this example is not to establish that relapsed/refractory evidence can be transported to frontline disease under this particular model. It is to demonstrate that **moving to a new disease setting creates a new evidence question**. The team must decide what it is comfortable carrying forward, what could plausibly change, and what additional evidence is needed before the new study is justified. The statistical bridge simply makes those assumptions visible enough to examine and challenge.

5.5 De-escalation: can treatment burden be reduced while preserving acceptable efficacy?

The fourth example asks a different kind of development question. Rather than asking whether the evidence is sufficient to expand, randomize, or move treatment into a new setting, it asks whether the evidence is sufficient to justify studying less treatment.

This situation arises when standard therapy is highly effective but treatment burden remains clinically important. The relevant question is then not simply whether a new regimen is active. It is whether treatment can be reduced while preserving an efficacy level that the clinical team would still consider acceptable. In E2D, that question is handled in the same way as the advancement examples: the action is defined first, the future study is specified, and success is defined in terms that match the clinical objective.

In the illustrative demonstration, the current evidence consists of 40 observations under standard treatment, with true efficacy probability 0.95. A Beta(1,1) prior is updated using the observed standard-treatment responses. For a prespecified efficacy loss d, the corresponding de-escalated efficacy probability is represented as

$$p_{\text{DE}} = p_{\text{STD}} - d$$

The value d is an assumed efficacy loss for the scenario, not an effect estimated from the 40 current standard-treatment observations.

The contemplated future study is a standalone randomized noninferiority trial of 120 patients, with 60 patients per arm (Piaggio et al., 2012). Current standard-treatment observations inform prediction but are not borrowed into the formal analysis of the future trial. Future efficacy success requires

$$\Pr(p_{\text{DE}} - p_{\text{STD}} > -0.05 \mid D_{\text{future}}) \geq 0.90$$

The future study is also assumed to achieve the intended reduction in treatment burden. In this demonstration, that burden criterion is fixed as satisfied rather than modeled as a separate outcome. Complete future success therefore requires both preservation of efficacy within the prespecified noninferiority margin and the intended burden reduction. Current evidence supports undertaking the de-escalation study when

$$\text{PP}_{\text{DE}} \geq 0.17$$

The numerical values in this example are illustrative and should not be interpreted as defining how much efficacy loss would be clinically acceptable in an actual treatment-reduction study. The purpose of the example is broader: E2D is action-specific, not advancement-specific. The same framework can be used when the clinically relevant next question is whether to expand development, generate comparative evidence, move into a different setting, or reduce treatment burden. What changes is the future study and the definition of success—not the underlying evidence-to-decision logic.

# 6. Calibrating Action-Specific Support Rules

The predictive-support thresholds used in E2D should not be interpreted as universal standards for how much statistical confidence is needed before a treatment can move forward. Each threshold belongs to a **specific development action** and is calibrated against the consequences of supporting that action when the evidence is insufficient or failing to support it when the evidence is worthwhile.

For each illustrative action, we therefore prespecified two types of calibration scenarios. **False advance** refers to supporting an action in a scenario defined in advance as insufficient for that action. **False discard** refers to failing to support the action in a scenario defined in advance as worthwhile. These are operating characteristics of the statistical support rule under the simulation model; they are not retrospective judgments that a real clinical decision was correct or incorrect.

The definition of “insufficient” and “worthwhile” was action-specific. For example, the larger single-arm expansion was calibrated against efficacy and toxicity settings appropriate to that decision, whereas the randomized actions were calibrated around clinically meaningful treatment differences together with acceptable experimental-arm toxicity. The de-escalation example used prespecified levels of efficacy loss. Scenarios falling between the prespecified insufficient and worthwhile regions were treated as gray-zone settings and were used descriptively rather than to redefine the support rules.

The maximum tolerated false-advance rates were 0.20 for larger same-setting expansion, 0.10 for same-setting randomized Phase II, and 0.05 for frontline randomized Phase II and de-escalation; the false-discard tolerance was 0.20 for each action using its prespecified denominator. These values were chosen for the present demonstration and should not be interpreted as NANT-wide or pediatric-oncology standards. Their role is to show that an action-specific support rule can be calibrated transparently rather than selected informally after observing the simulation results.

For the randomized actions, efficacy and safety thresholds were calibrated jointly so that both components had to satisfy their respective support requirements. After the support rules were frozen, independent higher-precision simulations were used to verify that the prespecified calibration criteria were still met. Verification assessed the frozen rules; it did not simply select new thresholds after seeing the higher-precision results. Detailed calibration anchors, denominators, threshold-search procedures, Monte Carlo adjustments, and tie-resolution rules are provided in the Technical Supplement.

Table 2 summarizes the future-study success definitions, the frozen support rules, and their verification performance. The main purpose of the table is not to establish universal numerical cutoffs, but to document that the support decisions used in the worked example were prespecified, calibrated, and reproducible.

**Table 2. Quantitative future-study definitions, frozen support rules, and verification performance**

| Action | Contemplated future study | Future-success event | Frozen support rule | Verification |
|---|---|---|---|---|
| Larger single-arm expansion | Add 20 patients in same population, setting, dose, and endpoint context; combine current and added data. | Posterior efficacy > 0.20 with probability ≥ 0.90 and posterior toxicity < 0.30 with probability ≥ 0.90. | PPLSA ≥ 0.20; Development Fork required. | Point FA 0.1620; point FD 0.1938; PASS |
| Same-setting randomized Phase II | Standalone RCT, N=60, 30 per arm; current patients excluded from future formal analysis. | Clinically meaningful efficacy contrast > 0.15 with posterior probability ≥ 0.90 and experimental-arm toxicity < 0.30 with posterior probability ≥ 0.90. | PPE ≥ 0.16 AND PPT ≥ 0.41; Development Fork required. | Point FA 0.0370; point FD 0.0832; robust FA 0.0489; robust FD 0.0920; PASS |
| Frontline randomized Phase II | Standalone frontline/earlier-line RCT, N=60, 30 per arm; reference transport λ=1, ηT=0. | Frontline efficacy contrast > 0.15 with posterior probability ≥ 0.90 and experimental-arm toxicity < 0.30 with posterior probability ≥ 0.90. | PPE ≥ 0.27 AND PPT ≥ 0.39; Development Fork required. | Point FA 0.0421; point FD 0.1647; robust FA 0.0466; robust FD 0.1730; PASS |
| De-escalation | Standalone randomized noninferiority trial, N=120, 60 per arm; current standard-treatment evidence n=40. | Noninferiority: pDE − pSTD > −0.05 with posterior probability ≥ 0.90; burden-reduction criterion fixed as satisfied. | PPDE ≥ 0.17; no Development Fork. | Point FA 0.0000; point FD 0.0000 in evaluated anchors; PASS |

FA denotes false advance; FD, false discard. These calibration quantities apply to prespecified finite anchor scenarios and should not be interpreted as universal clinical thresholds or uniform continuous-parameter error control.

# 7. Simulation Evaluation of the Decision Framework

The simulation study was designed to examine how the E2D support rules behave across clinically different evidence states, rather than to identify a single optimal development pathway.

For the three post-Fork advancement actions, the primary evaluation crossed current-trial efficacy probabilities of 0.15, 0.25, 0.40, and 0.55 with unacceptable-toxicity probabilities of 0.10 and 0.35, yielding eight scenarios. These settings were chosen to span weak, intermediate, and strong efficacy signals under more favorable and less favorable safety conditions. Within each simulated trial, the same observed current dataset was used to evaluate all three advancement actions in parallel after Development Fork passage: larger same-setting expansion, same-setting randomized Phase II, and frontline randomized Phase II. No hierarchy among the actions was imposed, and all combinations of support were allowed.

This structure was important because the primary question was not simply whether support increased or decreased as efficacy and toxicity changed. It was whether the **same current evidence could leave different sets of next studies open**. The simulation therefore retained the complete joint support pattern across the three actions, in addition to reporting the marginal probability of support for each individual action.

Broader action-specific evaluations were also conducted to examine how the randomized support rules behaved outside the primary eight-scenario grid. For the same-setting randomized Phase II action, efficacy probabilities of 0.15, 0.20, 0.25, 0.35, 0.40, and 0.55 were crossed with toxicity probabilities of 0.10, 0.20, 0.30, and 0.35, with control efficacy fixed at 0.20, giving 24 scenarios. For the frontline action, true frontline treatment differences of 0, $3/40$, $3/20$, and $5/22$were crossed with frontline experimental-arm toxicity probabilities of 0.10, 0.20, 0.30, and 0.35, giving 16 scenarios. These broader grids included insufficient, gray-zone, boundary, and worthwhile settings and were used descriptively outside the prespecified calibration anchors.

The standalone de-escalation example evaluated assumed efficacy losses of

$$d = 0, 0.02, 0.05, \text{ and } 0.10.$$

These scenarios were intended to show how the framework behaves under different treatment-reduction assumptions; they were not used to estimate a continuous clinical cutoff for acceptable efficacy loss.

Each formal scenario used 2,000 simulated current-trial replicates and, when prediction was required, 1,000 inner predictive simulations per action. Independent targeted verification of the frozen calibration rules used 10,000 outer replicates and 5,000 inner predictive simulations for the prespecified verification settings. Higher simulation precision was therefore concentrated on the calibration questions for which it was most consequential rather than applied uniformly to every descriptive scenario.

An important feature of the simulation design is that the fixed scenario efficacy and toxicity probabilities were used **only to generate the current trial data and define the evaluation scenario**. They were not supplied to the E2D predictive calculations. The predictive engines received the observed current data together with the prespecified future-study models, control benchmarks, and carry-forward assumptions. Thus, the downstream support decisions were based on what the simulated development team would have observed, not on privileged knowledge of the underlying generating truth.

The evaluation recorded Development Fork probability, support for each individual action, support conditional on reaching the Fork, and the complete pattern of which actions were supported together. Additional operating characteristics included action-specific false advance and false discard, randomized efficacy and safety component behavior, and Monte Carlo uncertainty. Detailed scenario definitions, denominators, numerical integration procedures, random-number generation, and complete operating-characteristic tables are provided in the Technical Supplement.

# 8. Results

## 8.1 Different evidence states leave different development paths open

The central quantitative result of the E2D demonstration was not simply that support increased with efficacy and decreased with toxicity. More importantly, different current evidence states led to different sets of supportable next studies.

For the three post-Fork advancement actions, each simulated trial could support additional same-setting single-arm expansion, same-setting randomized Phase II, frontline randomized Phase II, any combination of these actions, or none. Support was deliberately nonexclusive; E2D did not require the actions to follow a fixed hierarchy or force the evidence into a single development recommendation.

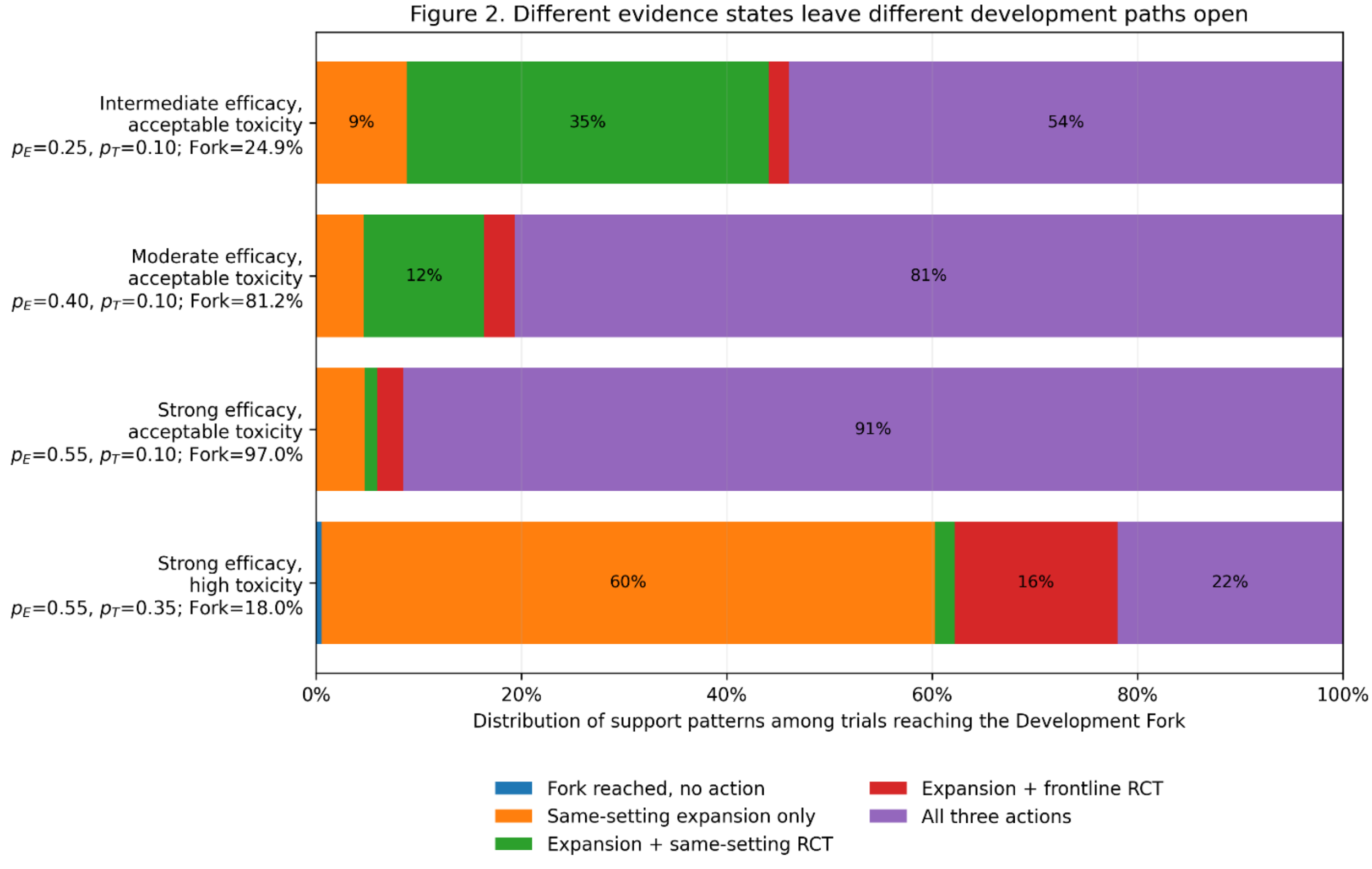

**Figure 2. Different evidence states leave different development paths open.**
Bars show the distribution of nonexclusive support patterns among simulated trials that reached the Development Fork for four representative primary scenarios. The three advancement actions are additional same-setting single-arm expansion, same-setting randomized Phase II, and frontline randomized Phase II. Percentages are conditional on Development Fork passage; the probability of reaching the Fork is shown for each scenario. Several actions may be supported simultaneously. The figure describes which development options remain evidentially open at the decision point; it does not rank the actions or imply that all supported actions should be pursued. Full unconditional results for all eight primary scenarios are provided in Table 3 and Supplementary Table S8.

Figure 2 summarizes these joint support patterns among trials that reached the Development Fork, while Table 3 provides the corresponding end-to-end results across all simulated current-trial pathways. The distinction is clinically useful: "additional same-setting experience only," "same-setting randomization but not frontline," and "all three paths are supportable" represent different development states rather than different values on one generic scale of promise. The complete unconditional support-pattern probabilities are provided in Supplementary Table S8.

The support architecture did not impose a structural ordering among actions. Some non-nested patterns occurred in the finite simulations, whereas others were not observed; absence of a pattern should not be interpreted as a logical constraint.

## 8.2 Intermediate evidence can support same-setting development without yet supporting movement to frontline

The action-specific value of E2D was particularly apparent when the current evidence was intermediate. At $p_E=0.25$ and $p_T=0.10$, 24.85% of simulated trials reached the Development Fork. Among all simulated trials, 8.75% supported both larger same-setting expansion and same-setting randomized Phase II while not supporting frontline randomized Phase II. This corresponds to approximately 35% of trials that reached the Fork.

Clinically, this represents a recognizable development state: the evidence may be sufficient to continue development in the setting in which the signal was observed—including moving to comparative evaluation—while still being insufficient to justify a move into frontline disease.

This distinction arises because the two randomized actions ask different questions. Same-setting randomization requires sufficient evidence for comparative evaluation in the current disease context. Frontline development additionally requires assumptions about which findings remain informative when the population, expected control outcome, treatment context, and benefit-risk balance change. The result therefore illustrates the central E2D principle that evidence can be sufficient for one next study without being sufficient for another.

## 8.3 Safety concerns can narrow the set of supportable next steps even when activity is strong

Strong activity did not automatically make the same downstream actions supportable when safety was less reassuring. At $p_E=0.55$ and $p_T=0.35$, 18.0% of simulated trials reached the Development Fork. Larger same-setting expansion alone was supported in 10.75% of all simulated trials, corresponding to approximately 60% of Fork-reaching trials. All three advancement actions were simultaneously supported in only 3.95% of all trials.

The important result is not simply that greater toxicity reduces support. Rather, safety concerns can narrow the set of development options that remain supportable. A strong activity signal may still leave the evidence sufficient for additional learning in the current setting while being insufficient for a larger comparative or context-changing commitment.

This scenario has an unfavorable toxicity truth and should not be interpreted as establishing that single-arm expansion is clinically preferable when toxicity is high. Its role is to demonstrate that different development actions can require different levels of safety reassurance.

## 8.4 Strong evidence can make several development paths supportable at the same time

When efficacy was stronger and toxicity remained low, the support set broadened substantially. At $p_E=0.40$ and $p_T=0.10$, all three advancement actions were supported simultaneously in 65.45% of simulated trials. At $p_E=0.55$ with the same toxicity probability, simultaneous support increased to 88.8%.

E2D does not interpret simultaneous support as ambiguity that must be eliminated statistically. Instead, it indicates that the same evidence base can make several scientifically different future studies defensible at the same time.

At that point, the role of the quantitative analysis is to make clear which options are supported and what that support depends on, not to manufacture a single winner. The ultimate choice among those options remains a multidisciplinary development decision.

## 8.5 De-escalation shows that E2D can support a different development objective

The de-escalation example addressed a qualitatively different question from the three post-Fork advancement actions: whether the evidence was sufficient to justify studying less treatment.

Under the illustrative noninferiority setup, de-escalation was supported in all 2,000 replicates when the assumed efficacy loss was 0 or 0.02 and in none of the 2,000 replicates when the assumed loss was 0.05 or 0.10. The predictive-probability distributions shifted accordingly across the frozen support threshold.

The observed transition between the evaluated losses of 0.02 and 0.05 should not be interpreted as a universal clinical cutoff. The more important result is conceptual: E2D is action-specific, not advancement-specific. The same framework can evaluate whether evidence is sufficient to gain more experience, randomize, move into a different disease setting, or reduce treatment burden by changing the future study and the definition of success.

## 8.6 Calibration and broader operating characteristics support the quantitative implementation

The support rules underlying these development patterns were prespecified and calibrated rather than chosen after examining the final simulation results. All four frozen rules passed independent higher-precision verification under their prespecified criteria. For larger same-setting expansion, the point false-advance and false-discard probabilities were 0.1620 and 0.1938. The corresponding values were 0.0370 and 0.0832 for same-setting randomized Phase II and 0.0421 and 0.1647 for frontline randomized Phase II. The de-escalation rule showed no observed false advances or false discards at the evaluated calibration anchors.

Broader same-setting and frontline grids showed graded changes in support across efficacy and toxicity settings, providing additional descriptive checks of the frozen rules. These expanded evaluations did not alter calibration, and formal error-control claims remain restricted to the prespecified calibration anchors. The former main-text heatmaps summarizing these broader operating characteristics are provided in the Technical Supplement rather than serving as a principal clinical result.

Taken together, the quantitative evaluation supports the intended role of the E2D engine: not to assign a single global strength-of-evidence score, but to distinguish among different sets of next studies that the current evidence is sufficient to support.

Figure 3. Different evidentiary domains can limit different randomized development actions.

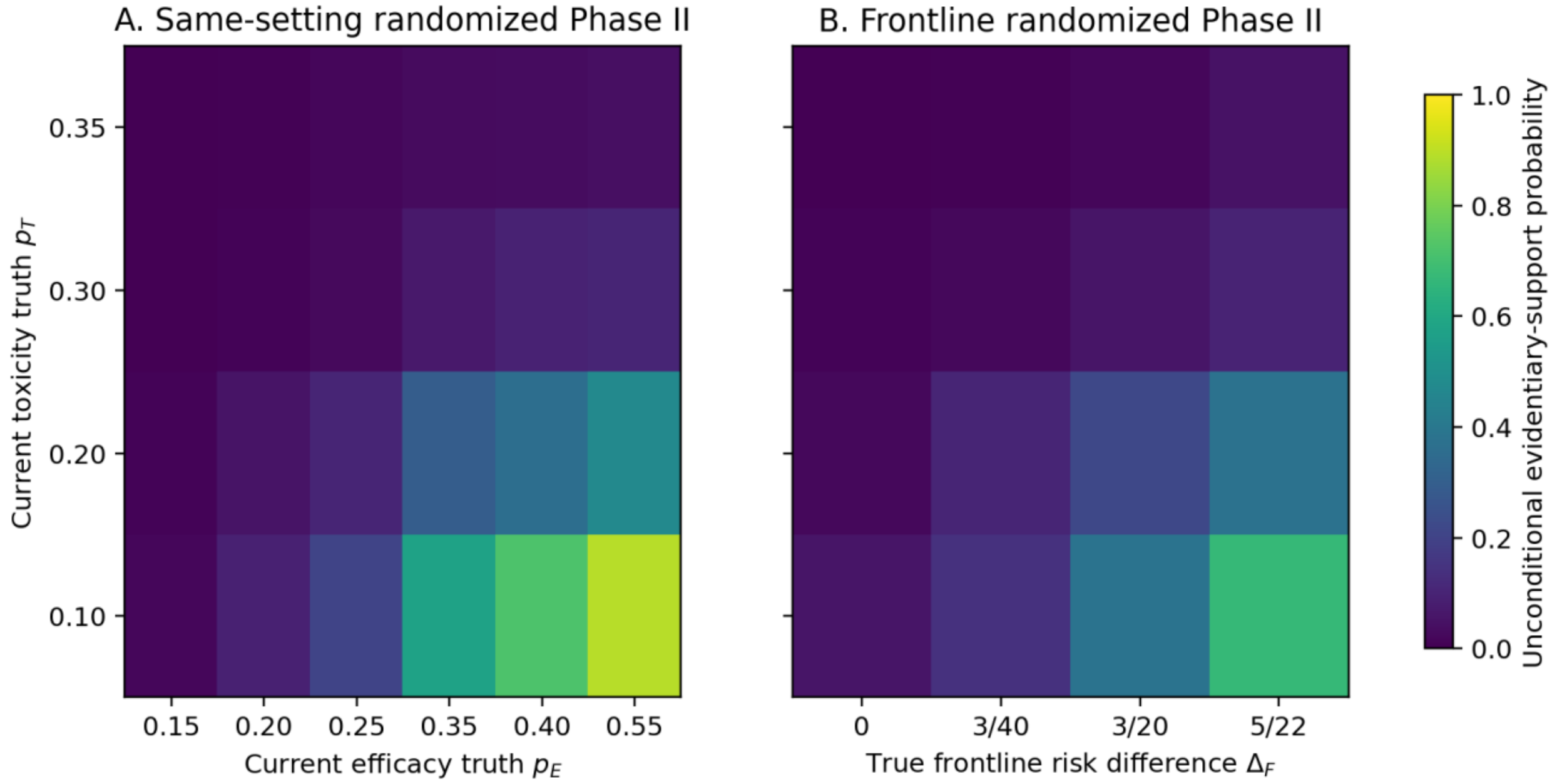

Heatmaps show unconditional end-to-end evidentiary-support probabilities for two contemplated randomized actions. Panel A shows same-setting randomized Phase II over current efficacy and toxicity truths. Panel B shows frontline randomized Phase II over true frontline risk differences and frontline experimental toxicity under the specified reference transport model. Only evaluated grid points are displayed. Full-region results are descriptive outside prespecified calibration anchors; formal error-control claims do not extend to the entire continuous parameter space. The frontline results do not empirically validate transportability.

# 9. Discussion

Early-phase oncology evidence is often summarized by asking whether a regimen is promising, whether a study is positive, or whether development should continue. Those judgments are useful shorthand, but they compress several clinically different decisions into a single label. An encouraging early signal may justify learning more in the same setting, initiating a randomized comparison, moving the regimen into an earlier-line population, modifying dose or schedule, focusing development on a biomarker-defined subgroup, reducing treatment burden, or first collecting additional evidence. These are not stronger or weaker versions of one generic decision. They are different next-study questions, with different evidence requirements, assumptions, benefit-risk considerations, and consequences of being wrong. E2D is intended to make those distinctions explicit by asking a more specific question: **what is the evidence available now sufficient to support next?**

The principal contribution of E2D is therefore not a new predictive-probability formula. Predictive modeling provides one quantitative layer within a broader clinical-development framework. The decision point and the candidate actions are first defined clinically. Each action is then tied to a specific future study, a clinically meaningful definition of success, and the evidence that would be needed before taking that step. Quantitative methods can subsequently characterize how strongly the current evidence supports that future study, how dependent that support is on key assumptions, and what important information remains missing. The clinical question comes first; the statistical model is used to make the evidentiary basis of that question more explicit and testable.

The numerical demonstration illustrates why this action-specific structure matters. The most informative result was not simply that support increased with efficacy and decreased with toxicity. Rather, the simulations produced **clinically different sets of supportable next studies**. Under intermediate efficacy with acceptable toxicity ($p_E, = 0.25\ p_T = 0.10$), 24.85% of simulated trials reached the Development Fork. Among those Fork-reaching trials, approximately 35% supported additional same-setting experience and same-setting randomized Phase II while not supporting movement to frontline disease. This represents a recognizable development state: the evidence may be mature enough to justify comparative evaluation in the setting in which the signal was observed, while still being insufficient to justify changing the disease setting and benefit-risk context. A single global "promising" label would not preserve that distinction.

Safety created a different pattern. In the stress scenario with strong efficacy but high unacceptable-toxicity probability ($p_E, = 0.55\ p_T = 0.35$), 18.0% of simulated trials reached the Development Fork, and approximately 60% of those Fork-reaching trials supported additional same-setting expansion without supporting either randomized action. The point is not that single-arm expansion should be preferred whenever toxicity is high; this scenario was intentionally unfavorable with respect to safety. Rather, it illustrates that strong activity does not automatically make the same downstream commitments supportable when the safety evidence is less reassuring. Different actions may require different degrees of safety reassurance, and safety concerns can narrow the set of defensible next steps before they eliminate every possible evidence-generating action.

At the other extreme, sufficiently compelling evidence can make several development paths supportable at the same time. At $p_E = 0.40$and $p_T = 0.10$, all three advancement actions were supported simultaneously in 65.45% of simulated trials; at $p_E = 0.55$with the same toxicity probability, that proportion increased to 88.8%. E2D does not interpret this as failure to identify a winner. The studies ask different scientific questions and may all be defensible from the same evidence base. Once several paths are supportable, the role of the quantitative analysis is to make those options and their assumptions visible, not to manufacture an artificial ranking among them.

The support architecture also does not impose a structural hierarchy among actions. The simulation allowed all combinations of the three post-Fork actions, and some non-nested support patterns occurred even though others were not observed in the finite simulations. Absence of a particular pattern should therefore not be interpreted as a logical constraint. This is also an appropriate target for clinical co-design: investigators may decide that some development pathways should be treated as logically or clinically dependent in a particular application, whereas others should remain independently evaluable.

Movement from relapsed/refractory disease into earlier-line or frontline therapy illustrates why changing the clinical setting requires more than stronger evidence of the same type. The current demonstration used an explicit statistical bridge linking relapsed/refractory and frontline efficacy and toxicity, but those relationships were intentionally illustrative rather than empirically validated. In practice, the relevant clinical question is more direct: **which findings from the current setting can reasonably inform the new setting, and which need to be established again?** Efficacy, durability, safety, exposure, disease biology, endpoint relevance, expected control outcomes, and tolerance for toxicity may not carry forward in the same way. The value of making these assumptions explicit is not that a statistical transport model can resolve them automatically; it is that the development team can see exactly where a proposed move depends on evidence versus judgment.

The de-escalation example extends the same logic in a different direction. E2D is not inherently an engine for escalating treatment intensity, trial size, or development commitment. In the illustrative de-escalation analysis, all 2,000 replicates supported the action when the assumed efficacy loss was 0 or 0.02, whereas none supported it when the assumed loss was 0.05 or 0.10. These evaluated losses should not be interpreted as defining a universal clinical cutoff. Their more important role is to demonstrate that the framework is **action-specific rather than advancement-specific**. When the relevant clinical question is whether treatment burden can be reduced while preserving acceptable efficacy, the future-success event changes accordingly, but the same evidence-to-decision architecture still applies.

An important implication is that the most useful E2D output may be richer than a support probability alone. A development team needs to know not only whether an action is currently supportable, but also **why it is or is not supportable**. Limited support may arise because efficacy follow-up is immature, safety experience is inadequate, the future comparator is uncertain, a biomarker is poorly characterized, or confidence in carrying evidence across settings is weak. These are clinically different problems because they imply different next steps. The question therefore becomes not simply "should we proceed?" but also "what information is preventing us from proceeding, and what additional evidence would actually change the decision?"

This creates a natural evidence-planning extension of E2D. If an important action is not yet supportable, the next development action may itself be to reduce the uncertainty that is blocking it. Depending on the question, that may mean enrolling additional patients, waiting for more mature follow-up, obtaining PK or exposure information, improving biomarker characterization, strengthening the control benchmark, or seeking relevant external evidence. The present paper does not formally optimize among those evidence-generating strategies. Its contribution is to create a structure in which the missing information is identifiable and can eventually become the target of a more formal evidence-acquisition analysis.

The quantitative rules used in the demonstration were calibrated rather than chosen ad hoc. All four frozen support rules passed their prespecified independent higher-precision verification criteria. Those results are important because they establish that the decision patterns described above arise from prespecified, reproducible support rules. At the same time, calibration is best viewed as validation of the quantitative engine rather than the primary clinical result. The central clinical question remains which future studies are supportable under the evidence and assumptions available at the decision point.

E2D also makes explicit the division of responsibility between clinical and statistical investigators. Clinical investigators are needed to define the real decision, identify the actions genuinely under consideration, determine what would constitute meaningful future success, specify the relevant benefit-risk context, and judge which assumptions about population, comparator, endpoint, biology, or disease setting are credible. Statistical investigators can formalize those questions, propagate uncertainty, calibrate support rules, evaluate sensitivity to assumptions, and characterize operating behavior. Neither role replaces the other. The framework is intended to provide a common interface between clinical judgment and quantitative evidence rather than an automated recommendation system.

For the same reason, evidentiary support should remain distinct from the final clinical-development choice. Several actions may be adequately supported while only one is feasible or strategically appropriate at a particular time. Patient availability, competing studies, drug supply, operational feasibility, regulatory strategy, clinical urgency, investigator judgment, and patient or family considerations may determine which supported action is ultimately pursued. These factors can be decisive without needing to be forced into a single statistical probability. E2D therefore structures the evidentiary component of the decision while preserving a separate multidisciplinary space for choosing among supportable options.

For a consortium such as NANT, this separation may provide an additional practical benefit: decision transparency and institutional memory. A structured E2D dossier could document what decision was being made, what evidence was actually available at that time, which actions were considered, what future study each action implied, what assumptions were required, what quantitative support was obtained, what evidence gaps remained, and which clinical or operational considerations ultimately shaped the choice. The goal would not be to make development decisions algorithmic, but to make the reasoning behind them easier to revisit, challenge, and learn from.

The present framework is intentionally modular. The quantitative examples in this paper—additional same-setting experience, same-setting randomized Phase II, movement to frontline disease, and de-escalation—illustrate different classes of development questions rather than a fixed menu. Other applications could include biomarker-enriched development, dose or schedule optimization, external or hybrid controls, platform-trial graduation, time-to-event outcomes, sequential evidence accumulation, or other treatment-optimization questions. In a real application, however, these actions should be included only when they represent genuine alternatives faced by the development team.

Several limitations define the scope of the current demonstration. The statistical implementation uses simplified binary efficacy and toxicity outcomes and a limited set of future-study designs. The candidate actions, success definitions, evidence requirements, support thresholds, and cross-setting assumptions were chosen for illustration and have not been established as NANT-wide clinical standards. The frontline example makes cross-setting assumptions explicit but does not empirically validate that relapsed/refractory efficacy or toxicity can be carried into frontline disease. The calibrated support thresholds belong to the prespecified demonstration and should not be interpreted as universal standards of evidence. Finally, real development decisions may depend on information that was not jointly modeled here, including response durability, marrow or MIBG response, dose and schedule, PK or exposure, cumulative toxicity, biomarkers, disease biology, and other disease-specific evidence.

These limitations point directly to the next stage of development. Moving from the present proof of principle to a clinically useful application requires more than making the simulation more elaborate. It requires a real decision point, clinically authentic candidate actions, clinically agreed definitions of future success, credible expectations for the comparator and future setting, explicit sensitivity analyses, and a record of what was actually known when the decision was made. Most importantly, it requires clinical co-design with investigators who make these decisions in practice.

Clinical co-design is therefore not a final endorsement step after the statistical framework has been completed. It is part of the framework itself. Investigators should be able to remove unrealistic actions, add missing ones, redefine what would count as future success, require additional evidence domains, reject assumptions about what can be carried across settings, or conclude that a particular development transition is not supportable from the available evidence. Such disagreement is useful: it exposes exactly where the decision depends on assumptions or missing information that might otherwise remain implicit.

The long-term opportunity for E2D is broader than a single predictive-probability method. Its role is to provide a common structure for stating **what we are considering doing next, what evidence we would want before doing it, what we know now, what we are assuming, and what remains missing**. Quantitative methods can then make those elements more explicit and auditable, while the multidisciplinary development team retains authority over the final choice. For rare pediatric oncology, where patient populations are small and each development decision carries substantial opportunity cost, that shared language may ultimately be more valuable than any single statistical score.

**Table 4. Proposed action-specific E2D evidence dossier for a real development discussion**

| Element | Question to document |
|---|---|
| Decision point | What decision are we making now, and what information was actually available at this time? |
| Next action under consideration | What specific next step are we genuinely considering? |
| Future study | If we take that step, what study would we actually do—population, comparator, endpoint, estimand, design, sample size, and safety requirements? |
| What would count as success? | What evidence from that future study would make us consider it successful? |
| Evidence we would want to review | Which efficacy, safety, durability, dose/exposure, biomarker, comparator, biological, or other information matters for this decision? |
| Evidence available now | Which of those pieces do we already have, and how mature or reliable are they? |
| What are we carrying forward? | What from the current setting are we using to inform the future study, what may change, and what should be established separately? |
| Quantitative support | How strongly does the current evidence support this future study? Are separate efficacy and safety components helpful? |
| What changes the answer? | Which assumptions or inputs materially change the level of support? |
| What is missing? | What information is preventing an important option from being supported now? |
| What would change our mind? | What additional feasible evidence could reduce that uncertainty enough to change the decision? |
| Multidisciplinary decision | Which supported option is ultimately pursued, and what clinical, operational, regulatory, or patient-centered considerations influenced that choice? |

# 10. Conclusion

E2D does not ask whether a treatment is promising in the abstract, nor does it attempt to identify a single statistically optimal development pathway. It asks a more clinically specific question:

**What is the evidence available now sufficient to support next—and what is it not yet sufficient to support?**

By organizing evidence around specific next-study decisions, E2D can distinguish among development states in which the appropriate next step may be to learn more in the current setting, proceed to comparative evaluation, consider movement into a new clinical setting, reduce treatment burden, or recognize that several development paths are simultaneously supportable.

Statistics provide the quantitative infrastructure for this process—formalizing the future study, propagating uncertainty, estimating predictive support, calibrating decision rules, and examining sensitivity to assumptions. But statistics do not define the clinical question or choose the final action. The actions under consideration, the evidence needed before taking them, the credibility of assumptions about what can be carried forward, and the final choice among supported options remain clinically defined and multidisciplinary.

E2D should therefore be viewed not as a finished decision rule, but as a **clinical-statistical co-design framework** for making explicit what is known, what is assumed, what is missing, and what evidence would be needed to take the next step with greater confidence.

# Clinical Co-Design Appendix A. NANT-Motivated Clinical Co-Design of the E2D Framework

*This appendix is clinical co-design material. It is distinct from the separate Full Technical Supplement, which contains the complete statistical specification and reproducibility details.*

**Clinical Co-Design Appendix A. NANT-Motivated Clinical Co-Design of the E2D Framework**

This appendix is intended to support clinical co-design rather than to attach a NANT label to a finished statistical model. The purpose is to give NANT investigators a practical way to change the framework itself: the development decisions that should be represented, the evidence that matters for each decision, what would count as successful future evidence, and which assumptions about carrying information forward are clinically reasonable or unacceptable.

In a rare pediatric oncology consortium, an encouraging early-phase signal may create several legitimate next-study questions at the same time. The team may consider gaining additional experience in the current setting, initiating a randomized comparison, moving the regimen earlier in therapy, focusing development on a biomarker-defined population, changing dose or schedule, reducing treatment burden, or deliberately collecting additional evidence before making a larger commitment. E2D is intended to make these choices and their evidence needs explicit rather than collapse them into a single promising/not-promising judgment.

## A.1 A practical NANT-motivated E2D workflow

1. **Define the decision point.** What decision is actually being made now?
2. **Record what is known now.** What efficacy, safety, follow-up, durability, dose/schedule, PK/exposure, biomarker, biological, external, or other information is actually available at this date?
3. **List the real next-step options.** Include only actions the clinical-development team is genuinely considering.
4. **Specify what each option would involve.** For each action, define the future study: population, comparator, endpoint, estimand, design, sample size, clinically meaningful effect, and safety requirements.
5. **Define what would count as success.**
6. **Identify what evidence would be needed before taking that step.**
7. **Compare those needs with what is available now.**
8. **Make clear what is being carried forward.** What from the current setting is assumed to remain informative, what may change, and what should be re-established?
9. **Quantify support and sensitivity** using an appropriate statistical approach.
10. **Identify what is missing.** What is preventing an important option from being supported now?
11. **Ask what would change the decision.** What additional feasible evidence could materially reduce that uncertainty?
12. **Record the final multidisciplinary choice separately**, including clinical, operational, regulatory, feasibility, and patient/family considerations.

## A.2 What is required before real-world use

Moving from the present proof of principle to real-world use requires more than replacing hypothetical probabilities with real numbers. It requires a real clinical decision point, authentic development options, clinically agreed definitions of what would count as successful future evidence, credible expectations for the comparator and future setting, and an explicit record of what information was actually available when the decision was made.

The current quantitative demonstration shows that an action-specific statistical layer can be implemented, calibrated, and evaluated reproducibly. It does **not** establish that the illustrative actions, success criteria, support thresholds, or assumptions about what can be carried from one disease setting to another are appropriate for a real NANT decision. Those elements should be refined through clinical co-design before prospective use.

## A.3 Proposed output for a real development meeting

A live E2D output should be a structured evidence dossier, not a ranked list of probabilities. For each development option, the dossier should show:

what study is being considered;

what would count as success;

what evidence the team would want before taking that step;

what evidence is available now;

what is being carried forward or assumed;

how strongly the current evidence supports the study;

what assumptions materially change that support;

what remains missing; and

what additional evidence could change the decision.

The final multidisciplinary decision should then be documented separately, together with the clinical, operational, regulatory, feasibility, or patient-centered considerations that influenced the choice.

## A.4 Suggested translational sequence

1. **Initial framework paper:** establish the E2D clinical architecture, demonstrate a transparent quantitative implementation, and provide a NANT-motivated co-design template.

2. **Clinical co-design:** work with NANT investigators to refine authentic development options, evidence needs, success definitions, and assumptions about what can reasonably be carried forward.
3. **Retrospective decision replay:** select a historical development decision, freeze the information that was actually available at that time, and apply E2D without allowing later outcomes to redefine the original problem.
4. **E2D Engine v0.1:** develop software that reproduces the quantitative demonstration and at least one real decision case, including sensitivity and audit outputs.
5. **Prospective pilot:** use an E2D evidence dossier during a real NANT or institutional development discussion and evaluate its interpretability, burden, and practical usefulness.
6. **Broader platform development:** after clinical co-design and proof of concept, extend the approach toward a reusable rare pediatric oncology evidence-to-decision system.

# Evidence-to-Decision (E2D) — Complete Technical Supplement

**Purpose of this technical supplement.** The companion clinical-framework manuscript places the clinical Evidence-to-Decision architecture in the foreground. This v2.2 supplement preserves the complete statistical specification underlying the numerical demonstration: current-trial screening, posterior predictive calculations, action-specific bridges, future-study analyses, calibration definitions, threshold search and selection, targeted verification, simulation grids, numerical integration, random-number generation, operating-characteristic tables, and supplementary figures. The statistical rules and frozen numerical results are unchanged by the clinical reframing or by the revised Results presentation.

**Interpretation.** These are demonstration-specific statistical choices, not universal NANT or pediatric-oncology standards. The main manuscript explains how clinically authentic actions, evidence requirements, success definitions, and transport assumptions should be co-designed before this quantitative layer is used in practice.

## Road map from the main manuscript to this technical supplement

This map is intended to let clinical readers skip technical detail while allowing statistical readers to reconstruct every quantitative component without ambiguity.

| Main manuscript topic | Technical supplement location |
|---|---|
| Current early-phase trial and Development Fork | S1; Supplementary Table S1 |
| General action-specific predictive probability | S2 |
| Larger single-arm expansion | S3; Supplementary Table S2 |
| Same-setting randomized Phase II | S4; S7–S13; Tables S2–S5; Fig S2A |
| Frontline / earlier-line randomized Phase II | S5; S7–S13; Tables S2–S4, S6; Fig S2B |
| De-escalation | S6; S7–S13; Tables S2–S4, S7; Fig S1 |
| Calibration, false advance / false discard, threshold selection | S7–S11; Supplementary Tables S3–S4 |
| Simulation grids, numerical computation, and operating characteristics | S12–S13; Tables S5–S8; Figs S1–S2 |
| Random-number generation and reproducibility | S14 |
| Interpretation boundaries | S15 |

| Joint post-Fork support-pattern outputs used by the revised main Results | S12.1; Table S8 (source for revised main Fig 2/Table 3) |
|---|---|

# S1. Current-trial statistical model and screening rules

## S1.1 Efficacy model

The current post-recommended phase II dose (post-RP2D) expansion uses a binary favorable efficacy outcome. Conditional on fixed scenario efficacy probability $p_E$, patient outcomes are independent Bernoulli variables and the response count among $n$ evaluable patients is binomial. The current efficacy prior is Beta(1,1), independent of the current toxicity prior. For response count $x_E$, the posterior is

$$p_E \mid x_E, n \sim \text{Beta}(1 + x_E, 1 + n - x_E).$$

This posterior represents uncertainty for prediction; current efficacy screening itself uses the prespecified Simon optimal two-stage design (Simon, 1989). Reference design inputs are null response probability $p_0 = 0.20$, alternative probability $p_1 = 0.40$, one-sided type I error at most 0.10, and power at least 0.80. For candidate $(n_1, r_1, N, r)$, Stage 1 continues when $X_{E,1} > r_1$, and final efficacy passes when $X_{E,1} + X_{E,2} > r$. The exact efficacy-only rejection probability at response probability $p$ is

$$R(p) = \sum_{x_1 = r_1 + 1}^{n_1} \text{Bin}\,(x_1; n_1, p) P\{\text{Binomial}(N - n_1, p) > r - x_1\},$$

where $\text{Bin}(x; n, p)$ denotes a binomial probability mass. Expected sample size under the null is

$$EN(p_0) = n_1 + (N - n_1) P_{p_0}\big(X_{E,1} > r_1\big).$$

An exact integer search minimizes this expectation subject to $R(p_0) \le 0.10$ and $R(p_1) \ge 0.80$. The configured search permits total sample size up to 200. Numerical comparisons use tolerance $10^{-12}$. Equal optimal expectations are resolved by smaller $N$, then smaller $n_1$, $r_1$, and $r$, in that order. An infeasible search is reported rather than relaxing the design constraints.

The frozen selected design is $n_1 = 12, r_1 = 2, N = 25$, and $r = 7$. Thus, Stage 1 efficacy continuation requires at least three responses among 12, and final efficacy passage requires at least eight responses among 25. There is no additional Bayesian current-trial efficacy gate. Simon error and expected-sample-size quantities refer to efficacy-only screening; the actual current pathway additionally requires safety passage.

## S1.2 Safety model

Unacceptable toxicity is binary. Conditional on fixed toxicity probability $p_T$, its count is binomial; efficacy and toxicity are independent conditional on their fixed scenario probabilities. With independent Beta(1,1) prior, the toxicity posterior is Beta$(1 + x_T, 1 + n - x_T)$. The maximum acceptable toxicity probability is $T_{\max} = 0.30$.

At interim review, toxicity stopping occurs when

$$P\big(p_T > 0.30 \mid x_{T,1}, 12\big) \ge q_{\text{stop}} = 0.655.$$

This is equivalent to at least five toxicity events among 12 patients. At final review, safety passage requires

$$P(p_T < 0.30 \mid x_T, 25) \ge q_{\text{pass}} = 0.540,$$

equivalent to no more than six events among 25. Posterior probabilities are computed from the Beta cumulative distribution function; equality at the posterior threshold counts as stopping or passage, respectively.

Safety thresholds were calibrated in a standalone safety-only sequential process, without conditioning on efficacy. The acceptable and unacceptable toxicity anchors are 0.20 and 0.40. Write $A(x_1; q_s)$ for the indicator that the interim posterior exceedance probability is strictly below $q_s$, and $B\big(x_1 + x_2; q_f\big)$ for the indicator that the final posterior probability below 0.30 is at least $q_f$. Exact sequential safety passage is

$$Q(p_T; q_s, q_f) = \sum_{x_1=0}^{12} \sum_{x_2=0}^{13} \text{Bin}\,(x_1; 12, p_T)\text{Bin}(x_2; 13, p_T)A(x_1; q_s)B(x_1 + x_2; q_f).$$

The complementary interim stopping probability is obtained by summing the Stage-1 binomial masses for counts failing $A$. Calibration searches posterior thresholds from 0.500 through 0.999 in increments of 0.001. First, among interim thresholds limiting stopping at $p_T = 0.20$ to at most 0.10, it maximizes stopping at $p_T = 0.40$. With that interim rule fixed, final calibration limits $Q(0.40; q_s, q_f)$ to at most 0.10 and maximizes $Q(0.20; q_s, q_f)$. Ties select the smallest applicable posterior threshold. The resulting count boundaries and thresholds above remain fixed throughout downstream evaluation.

This enumeration includes only paths that survive interim safety review. A final-only binomial tail is therefore not substituted for the overall sequential safety-passage probability. In the actual trial, efficacy futility can further terminate enrollment, but does not redefine the standalone safety calibration.

## S1.3 Stage-1 pause and Development Fork

Enrollment stops temporarily after the twelfth Stage-1 patient. Review waits until all 12 patients have completed both efficacy and toxicity assessments. There is no accrual during this pause, no accrual overrun, and no continuation based on partially observed outcomes. Interim efficacy and safety failures are assessed simultaneously. Their separate indicators are retained even when both occur; safety precedence applies only to a mutually exclusive stop-reason label.

Continuation requires both interim gates to permit enrollment. A continuing trial enrolls 13 further patients, reaching 25, and final review again awaits complete efficacy and toxicity assessment. Final failures are retained separately, with safety precedence only for mutually exclusive labels. If $F$ denotes Fork passage,

$$F = I(\text{interim continuation})I(X_E \geq 8)I(X_T \leq 6).$$

Every trial with $F = 0$ receives zero support for all three post-RP2D actions. De-escalation is standalone and does not use this screening pathway.

Accrual follows a Poisson process, equivalently independent exponential interarrival times. The reference rate is 12 patients/year, converted to $12/365.25$ patients/day. Toxicity and efficacy assessment delays are fixed at 28 and 56 days after enrollment. Consequently, interim review occurs 56 days after the last Stage-1 enrollment. If continuation is allowed, new interarrival times begin from that review time; final review occurs 56 days after the final enrollment. Current-decision time is measured from expansion initiation to the applicable completed review and next-action evaluation. Statistical computation adds no calendar time, and conducting a contemplated future study is excluded. Complete evaluability is assumed: there is no missingness, dropout, replacement, imputation, or pending final outcome. Current sample size is therefore exactly 12 or 25. The standalone de-escalation example starts with pre-existing evidence and has no simulated current-accrual time.

# S2. General predictive-probability computation

For observed current data $D$ and an action-specific future-success event $S_a$, define

$$PP_a(D) = P(S_a = 1 \mid D).$$

This is a predictive probability conditional on the contemplated future design and its specified uncertainty model, not an unconditional probability that a treatment is generally promising (Spiegelhalter et al., 1986). Computation uses five steps. First, observed current counts update the relevant Beta posteriors. Second, latent future-generating probabilities are drawn from these posteriors and any prespecified uncertain benchmarks. Third, action-specific bridges map those draws into the contemplated future setting. Fourth, a future dataset is generated and analyzed under its separate formal-analysis model. Fifth, the applicable future-success indicators are evaluated.

The Beta-binomial predictive construction used in this demonstration is also consistent with established predictive-probability approaches in Phase II oncology trials (Lee and Liu, 2008).

For $M_{\text{inner}}$ inner simulations,

$$\widehat{PP}_a(D) = \frac{1}{M_{\text{inner}}} \sum_{m=1}^{M_{\text{inner}}} I\left(S_a^{(m)} = 1\right).$$

Randomized component probabilities use the same expression for efficacy and safety indicators separately. All future counts within one inner replicate belong to one paired future-study simulation. Component and optional joint summaries are calculated from those same replicates, without an additional posterior Monte Carlo layer.

Fixed outer truths generate current observations and determine scenario classifications; they are never supplied directly to the predictive engines. The decision model receives observed counts and the prespecified posterior, benchmark, design, and bridge inputs. A fixed outer control truth and an uncertain predictive control benchmark remain distinct. Outer truth is not redrawn from the benchmark distribution within each outer replicate. Likewise, a prespecified bridge parameter is a model input, not privileged knowledge of an otherwise unobserved treatment probability.

# S3. Larger single-arm expansion

The larger single-arm expansion (LSA) action adds $m_{\text{add}} = 20$ patients in the same population, disease setting, dose, and efficacy/toxicity endpoint context. There is no between-cohort drift. Each independent current efficacy and toxicity posterior draw is carried forward unchanged by the identity bridge.

In inner replicate $m$, added efficacy and toxicity counts are generated as binomial variables with size 20 and their respective posterior-drawn probabilities. They are combined with the observed current counts, giving combined sample size $n + 20$ for each endpoint. The combined posteriors retain the original Beta(1,1) priors, adding all current and new successes and failures once. The current posterior is not treated as additional observations on top of the already included current data.

Complete future success requires

$$P(p_E > 0.20 \mid D_{\text{combined}}) \geq 0.90 \quad \text{and} \quad P(p_T < 0.30 \mid D_{\text{combined}}) \geq 0.90.$$

The scalar $PP_{\text{LSA}}$ is the probability that both combined-data conditions pass. Unlike the standalone randomized actions, current patients are included in this action's formal combined analysis. Count-indexed success evaluation depends on the observed current counts and the possible added counts.

Current evidentiary support is

$$I_{\text{support,LSA}} = F\,I\big(\widehat{PP}_{\text{LSA}} \geq 0.20\big).$$

The future posterior threshold 0.90 defines successful combined evidence. The current predictive threshold 0.20 defines evidentiary support for this action. Neither substitutes for the other, and the support rule does not imply that future success is certain.

# S4. Same-setting randomized Phase II

The contemplated future randomized study is a standalone randomized controlled trial (RCT), with total $N_{\text{RCT}} = 60$ and fixed 1:1 allocation: 30 experimental and 30 control patients. Current expansion data inform prediction but are not included in its formal future analysis.

## S4.1 Predictive model

Experimental efficacy and toxicity probabilities are independent draws from their respective current posteriors and are passed unchanged through identity bridges. Future control efficacy is independently drawn from Beta(2,8), with mean 0.20 and effective sample size 10. In general, benchmark shapes are $\mu_C ESS_C$ and $(1 - \mu_C)ESS_C$.

This benchmark represents uncertainty about future control performance. It is not used as an informative future-analysis prior. Benchmark and current posterior draws are mutually independent under the specified marginal models. Conditional on an inner draw, experimental responses, control responses, and experimental toxicity events are generated as separate binomial counts of size 30. Control-arm toxicity is neither simulated nor used to determine success.

## S4.2 Formal future analysis

Future formal analysis uses independent neutral Beta(1,1) priors for experimental efficacy, control efficacy, and experimental-arm toxicity. Only future observations update these priors. Future efficacy and safety criteria are

$$P(p_E - p_C > 0.15 \mid D_{\text{future}}) \geq 0.90 \quad \text{and} \quad P\big(p_{T,E} < 0.30 \mid D_{\text{future},T,E}\big) \geq 0.90.$$

The risk-difference event uses a strict difference greater than 0.15, whereas the posterior-probability comparison uses at least 0.90. Safety concerns experimental-arm toxicity alone. Complete future-study success is the conjunction of these two criteria. Posterior risk-difference probabilities are evaluated deterministically as described in S13.1.

## S4.3 Predictive support

Let $E_m$ and $T_m$ indicate efficacy and safety passage in paired inner replicate $m$. The production predictive summaries are the separate means of these indicators. Their optional joint diagnostic is the mean of $E_m T_m$, not a substituted product of estimated marginal probabilities.

The support rule is

$$I_{\text{support,RCT}} = F\, I\big(\widehat{PP}_{E,\text{RCT}} \geq 0.16\big) I\big(\widehat{PP}_{T,\text{RCT}} \geq 0.41\big).$$

Both component gates must pass; stronger efficacy cannot compensate for a failed safety component. The diagnostic joint probability is not thresholded for support or used as the calibration variable. Enabling its calculation does not require new random draws or alter the component estimates. The paired calculation preserves the simulated future-study relationship among the component indicators without imposing a different joint decision rule.

# S5. Frontline randomized Phase II

The frontline action contemplates a new standalone randomized trial of 60 patients, allocated 30 per arm. Current relapsed/refractory (R/R) patients inform prediction through explicit bridges but are not borrowed into the future formal analysis.

## S5.1 Control benchmarks

The uncertain R/R control efficacy benchmark is Beta(2,8), with mean 0.20 and effective sample size 10. The uncertain frontline control benchmark is Beta(5,5), with mean 0.50 and effective sample size 10. These benchmark draws are independent of one another and of both current experimental posterior draws.

The R/R benchmark defines a relative treatment effect for transport; the frontline benchmark anchors future control performance. They have different roles and are not replaced by their means during predictive calculation. Fixed control truths used to construct outer scenarios do not replace these uncertainty distributions.

## S5.2 Efficacy transport bridge

For each paired draw, define

$$\delta_R = \text{logit}(p_{ER}) - \text{logit}(p_{CR}).$$

$$\delta_F = \lambda \delta_R.$$

$$\text{logit}(p_{EF}) = \text{logit}(p_{CF}) + \delta_F.$$

The reference $\lambda = 1$ preserves the R/R log-odds treatment effect. This does not preserve an absolute risk difference across settings, because the baseline control probability changes. Future experimental efficacy is obtained by the inverse-logit transformation of the transported expression. The uncertainty in both control benchmarks and current experimental efficacy is propagated through each inner draw.

This is a prespecified development assumption, not an empirically established relationship between R/R and frontline disease. The simulation evaluates decisions under that assumption; it does not validate transportability.

## S5.3 Toxicity bridge

Current R/R toxicity is transported using

$$\text{logit}\big(p_{T,F}\big) = \text{logit}\big(p_{T,R}\big) + \eta_T.$$

The reference $\eta_T = 0$ preserves toxicity odds and hence the toxicity probability in each draw. The mapped experimental-arm probability is used to generate future toxicity events. No frontline control-arm toxicity outcome is generated or enters the safety component.

## S5.4 Future analysis and support

Each inner replicate generates frontline experimental and control response counts and experimental toxicity counts, all with size 30. Formal analysis uses neutral independent Beta(1,1) priors and future data only. The criteria remain

$$P(p_{EF} - p_{CF} > 0.15 \mid D_{\text{future}}) \geq 0.90 \quad \text{and} \quad P\left(p_{T,F} < 0.30 \mid D_{\text{future},T,F}\right) \geq 0.90.$$

Component predictive probabilities are estimated from the same paired simulations. Support requires

$$I_{\text{support,frontline}} = F\, I\left(\widehat{PP}_{E,\text{frontline}} \geq 0.27\right) I\left(\widehat{PP}_{T,\text{frontline}} \geq 0.39\right).$$

A joint future-success predictive probability remains diagnostic only. Neither the future posterior criteria nor the component gate adopts an alternative success architecture; the transport model changes future-generating probabilities, not the formal-analysis definitions.

# S6. De-escalation demonstration

The standalone demonstration begins with 40 standard-treatment observations generated under fixed standard efficacy truth 0.95. If $x_{\text{STD}}$ responses are observed, the current posterior is Beta$\left(1 + x_{\text{STD}}, 41 - x_{\text{STD}}\right)$. Prediction uses this observed-data posterior, not the generating truth.

Prespecified efficacy losses are $d_{\text{DE}} \in \{0, 0.02, 0.05, 0.10\}$. The loss is a fixed bridge assumption for each evaluated scenario and is not estimated from standard-treatment observations. For each current standard posterior draw, the mapped de-escalated probability is

$$p_{\text{DE}} = \min\{1, \max\{0, p_{\text{STD}} - d_{\text{DE}}\}\}.$$

Truncation to $[0,1]$ occurs before outcome generation. It is an implementation safeguard for a valid probability, not an additional clinical assumption or an independently estimated treatment effect.

The future noninferiority study is standalone with total $N_{\text{DE}} = 120$, allocated 60 per arm. Future standard and de-escalated response counts are generated conditionally on the paired bridged probabilities. Their shared standard posterior draw is retained through the bridge; a separate independent standard predictive draw is not introduced for the other arm. (Piaggio et al., 2012)

Future formal analysis uses neutral independent Beta(1,1) priors updated only with future arm-specific counts. Current standard observations are not included or borrowed. Efficacy success requires

$$P(p_{\text{DE}} - p_{\text{STD}} > -0.05 \mid D_{\text{future}}) \geq 0.90.$$

Treatment-burden reduction is a prespecified design attribute fixed as satisfied, not a simulated endpoint. Complete future success requires both noninferiority and that attribute. Thus $PP_{\text{DE}}$ is estimated from complete-success indicators, and support requires $\widehat{PP}_{\text{DE}} \geq 0.17$, without a Development Fork. The future posterior threshold 0.90 and current support threshold 0.17 have separate roles.

# S7. Calibration framework

Clinical truth regions organize evaluated scenarios into insufficient, gray-zone, and worthwhile classes. Formal calibration sets impose error constraints at prespecified finite scenarios within the insufficient and worthwhile classes. A boundary point retains its clinical classification; "boundary" is not a fourth class.

LSA truth is insufficient when $p_E \leq 0.20$ or $p_T \geq 0.30$, and worthwhile when $p_E \geq 0.40$ and $p_T \leq 0.20$. For either randomized action, truth is insufficient when its experimental-control risk difference is at most zero or experimental toxicity is at least 0.30; it is worthwhile when the difference is at least 0.15 and toxicity is at most 0.20. Frontline classification uses induced frontline parameters, not bridge coefficients directly. Remaining truths are gray-zone. De-escalation is insufficient for loss at least 0.05 or an unsatisfied burden gate, worthwhile for loss at most 0.02 with that gate satisfied, and otherwise gray-zone.

## S7.1 Error definitions

False advance means supporting an action under an insufficient truth; false discard means not supporting it under a worthwhile truth. All worst-case quantities are maxima over the applicable finite calibration sets, not scenario averages.

For randomized action $a$,

$$FA_a(c_E, c_T) = \max_{\theta \in \mathcal{A}_{0,a}} P_\theta\{F = 1, PP_{E,a} \geq c_E, PP_{T,a} \geq c_T\},$$

and

$$FD_a^{\text{Fork}}(c_E, c_T) = \max_{\theta \in \mathcal{A}_{1,a}} P_\theta\{PP_{E,a} < c_E \text{ or } PP_{T,a} < c_T \mid F = 1\}.$$

False advance is unconditional and end-to-end. A non-Fork replicate contributes no support. Randomized false discard is conditional on the Fork: failure of the current screen is not a downstream randomized-rule false discard. This distinction does not remove screening from the actual support pathway.

LSA retains end-to-end scalar errors. Its worthwhile-scenario false discard includes both Fork failure and subsequent failure of the scalar gate. De-escalation errors concern its standalone scalar support event. No Fork conditioning is applied to de-escalation.

## S7.2 Error tolerances

Maximum false-advance tolerances are 0.20 for LSA, 0.10 for same-setting randomized Phase II, 0.05 for frontline randomized Phase II, and 0.05 for de-escalation. Maximum false discard is 0.20 for every action, using its action-specific denominator. These are illustrative demonstration targets, not universal clinical standards. Randomized anchor constraints do not constitute guarantees across the full clinically classified or continuous parameter region.

# S8. Formal calibration anchors

## S8.1 Same-setting randomized Phase II

Control truth is fixed at $p_C^{\text{true}} = 0.20$. In ordered pairs $(p_E^{\text{true}}, p_T^{\text{true}})$, the insufficient anchors are

$$\mathcal{A}_{0,\text{RCT}} = \{(0.15,0.10), (0.15,0.35), (0.25,0.35), (0.40,0.35), (0.55,0.35)\}.$$

The worthwhile anchors are

$$\mathcal{A}_{1,\text{RCT}} = \{(0.40,0.10), (0.55,0.10)\}.$$

The remaining original scenario $(0.25,0.10)$ is gray-zone and is not a calibration anchor. These finite sets, rather than all worthwhile and insufficient points in the expanded evaluation grid, define formal randomized calibration.

## S8.2 Frontline randomized Phase II

Reference control truths are 0.20 in R/R and 0.50 in frontline, with $\lambda = 1$ and $\eta_T = 0$. Ordered pairs now denote true frontline risk difference and frontline experimental toxicity, $(\Delta_F, p_{T,F}^{\text{true}})$. The insufficient and worthwhile sets are

$$\mathcal{A}_{0,\text{frontline}} = \{(0,0.10), (0,0.35), (5/22,0.35)\}, \qquad \mathcal{A}_{1,\text{frontline}} = \{(5/22,0.10)\}.$$

The effect $5/22$ is retained as an exact rational quantity, not replaced by a rounded stored decimal. With the reference bridge it corresponds to R/R experimental efficacy truth 0.40. Anchor inclusion does not change the underlying clinical truth-region definition.

## S8.3 Larger single-arm expansion

LSA calibration uses the prespecified original eight-scenario post-RP2D grid. Its exact insufficient calibration set, in efficacy–toxicity pairs, is

$$\mathcal{G}_{0,\text{LSA}} = \{(0.15,0.10), (0.15,0.35), (0.25,0.35), (0.40,0.35), (0.55,0.35)\}.$$

Its exact worthwhile set is

$$\mathcal{G}_{1,\text{LSA}} = \{(0.40,0.10), (0.55,0.10)\}.$$

The scenario $(0.25,0.10)$ remains gray-zone and is evaluated but does not enter either worst-case calibration error. The listed insufficient and worthwhile scenarios are the prespecified LSA calibration anchors; the gray-zone scenario does not enter formal calibration.

## S8.4 De-escalation

The evaluated worthwhile calibration losses are exactly $d_{\mathrm{DE}} = 0$ and 0.02; insufficient losses are exactly 0.05 and 0.10. The burden gate is satisfied throughout this demonstration. These are finite evaluated anchors within the prescribed clinical regions, not a calibration of a continuously estimated loss cutoff.

# S9. Threshold search and selector

Scalar support thresholds are evaluated on $\{0.01, 0.02, \ldots, 0.99\}$. Randomized actions evaluate its full two-dimensional Cartesian product, with 99 efficacy thresholds and 99 safety thresholds. There is no lower bound of 0.50 and no finer local refinement. Each frontier applies candidate gates to scenario-specific predictive outputs; a new future-study simulation is not required for every threshold candidate.

Scalar feasibility requires both empirical worst-case point errors to meet their tolerances, after any required targeted higher-precision replacement. Randomized feasibility first requires evaluable worthwhile-anchor conditional errors and point-estimate compliance. Its robust conditions, using Monte Carlo standard error (MCSE), are

$$FA_{a,\mathrm{robust}} = \max_{\theta \in \mathcal{A}_{0,a}} \left[ \widehat{FA}_{a,\theta} + 2\,MCSE_{FA,a,\theta} \right] \le \alpha_a^{FA},$$

and

$$FD_{a,\mathrm{robust}}^{\mathrm{Fork}} = \max_{\theta \in \mathcal{A}_{1,a}} \left[ \widehat{FD}_{a,\theta}^{\mathrm{Fork}} + 2\,MCSE_{FD,a,\theta} \right] \le \beta_a^{FD}.$$

Each anchor-specific error is adjusted before taking its maximum. This is not the maximum point estimate plus an MCSE attached to one selected worst anchor. Adjusted values are not clamped to one, and co-attaining anchors are not averaged or selected according to their smallest MCSE. Every anchor attaining a point or adjusted maximum is retained with its applicable estimate and MCSE.

If a worthwhile randomized anchor has no Fork-reaching trials, conditional false discard is NA. Targeted verification must resolve its denominator before feasibility, scores, or selection are computed. If it remains zero at maximum approved precision, calibration is considered not evaluable and no threshold pair is selected. The anchor is neither excluded nor imputed. For otherwise evaluable calibration, if no candidate satisfies the prespecified constraints, no evidentiary-support threshold is selected; an approximately feasible choice is not substituted.

## S9.1 Selector hierarchy

The common selector operates only on the final feasible scalar set or robust anchor-feasible randomized set. Its frozen point-error algebra is

$$M_{FA} = \frac{\alpha_a^{FA} - FA_a}{\alpha_a^{FA}}, \qquad M_{FD} = \frac{\beta_a^{FD} - FD_a}{\beta_a^{FD}},$$

and

$$S = \min(M_{FA}, M_{FD}).$$

Here, errors are full-precision formal worst-case point estimates, with randomized false discard using its Fork-conditional definition. MCSE-adjusted errors determine randomized candidate membership, but do not enter $S$. Retain all candidates within $10^{-12}$ of the maximum score.

Among these maximum-score candidates only, candidate A formally dominates candidate B when

$$FA_A \le FA_B, \qquad FD_A \le FD_B,$$

with at least one improvement beyond $10^{-12}$. Dominated candidates are removed. Dominance uses formal point errors, not threshold coordinates, MCSE, or descriptive boundary or gray-zone performance.

Remaining candidates share an exact formal-performance class only when worst-case point false advance, false discard, and $S$ agree within $10^{-12}$. For randomized actions, complete point-error co-attaining anchor sets must also be structurally identical. Rounded displayed values cannot establish a tie. If more than one nondominated exact class remains, the primary tie is unresolved and no threshold is selected; stringency cannot resolve different formal error tradeoffs.

A unique remaining candidate is selected directly. Within one exact class, duplicate-equivalent scalar candidates are collapsed and the largest $c$ is chosen. Randomized stringency instead uses the componentwise partial order:

$$c_E^{(1)} \geq c_E^{(2)}, \qquad c_T^{(1)} \geq c_T^{(2)},$$

with at least one strict inequality. A unique componentwise maximal pair is selected. If multiple incomparable maxima remain, the secondary tie is unresolved and no threshold pair is selected, with no tertiary rule and no preference for efficacy or safety. Coordinate sums, distances, visual choices, and full-region performance do not resolve ties. This hierarchy selects support thresholds within an action; it does not rank clinical-development actions.

# S10. Frozen evidentiary-support thresholds

The adopted current-evidence thresholds are

$$c_{\mathrm{LSA}} = 0.20, \qquad \left(c_{E,\mathrm{RCT}}, c_{T,\mathrm{RCT}}\right) = (0.16, 0.41),$$

and

$$\left(c_{E,\mathrm{frontline}}, c_{T,\mathrm{frontline}}\right) = (0.27, 0.39), \qquad c_{\mathrm{DE}} = 0.17.$$

These values are fixed for downstream evaluation. They are distinct from the future posterior success thresholds of 0.90. No threshold is selected anew for a full-region scenario or altered according to its observed predictive distribution. The same randomized component pairs are used in both the integrated three-action and respective descriptive full-region evaluations.

# S11. Targeted precision replacement and independent verification

## S11.1 Calibration-stage targeted precision replacement

Before an evidentiary-support rule is frozen, a calibration conclusion within $2 \times MCSE$ of its prescribed false-advance or false-discard tolerance triggers higher-precision evaluation of the relevant worst-case scenario or scenarios. During calibration or reproduction of the pre-freeze selection procedure, verified scenario-specific estimates replace the corresponding lower-precision estimates across the candidate frontier. The same final estimate set is used for feasible membership, point maxima, co-attainer sets, balanced scores, and tie classification; selectively replacing only a favorable candidate is not permitted.

## S11.2 Post-freeze independent targeted verification

After threshold selection and freezing, independent targeted verification assesses the frozen support rule. It uses 10,000 outer replicates and 5,000 inner predictive replicates, rather than uniformly increasing all formal scenarios. These are the maximum approved targeted counts.

The six targeted settings are LSA at $(p_E, p_T) = (0.40, 0.10)$; same-setting randomized Phase II at $(0.40, 0.10)$ and $(0.55, 0.35)$; frontline randomized Phase II at $\left(\Delta_F, p_{T,F}\right) = (0, 0.10)$ and $(5/22, 0.10)$; and de-escalation at loss 0.02 with standard efficacy truth 0.95. Corresponding fixed thresholds are those in S10. Each target retains the original design, priors, bridges, success criteria, and action-specific error denominator.

Verification assesses the frozen choice against its prespecified criteria, including point constraints and randomized robust constraints where applicable. It does not reopen threshold optimization merely because a diagnostic empirical optimizer differs. A frozen choice that failed its criterion would trigger scientific adjudication, not automatic recalibration or replacement. Reconstruction of a verification frontier therefore checks the procedure and quantifies its behavior without authorizing a new scientific rule.

# S12. Simulation scenario grids

## S12.1 Primary post-RP2D grid

The primary grid crosses $p_E^{\mathrm{true}} \in \{0.15, 0.25, 0.40, 0.55\}$ with $p_T^{\mathrm{true}} \in \{0.10, 0.35\}$, giving eight scenarios. Each replicate generates one complete current pathway. On reaching the Fork, that same observed current dataset informs all three advancement actions, evaluated in parallel with their respective predictive models and independent action streams.

The support vector uses fixed order LSA–same-setting RCT–frontline RCT. All eight three-bit patterns are permitted, including no support and simultaneous support. No implication among bits is imposed. Non-Fork trials contribute the all-zero vector, while Fork-reaching trials can also have all-zero support if all action gates fail. Marginal support, complete pattern probabilities, and conditional-on-Fork quantities distinguish these possibilities without ranking actions. Complete pattern probabilities are reported in Supplementary Table S8 and provide the source for the revised main-manuscript support-set display.

## S12.2 Same-setting full-region grid

The descriptive same-setting grid crosses efficacy truths $\{0.15, 0.20, 0.25, 0.35, 0.40, 0.55\}$ with toxicity truths $\{0.10, 0.20, 0.30, 0.35\}$, giving 24 scenarios, with control truth fixed at 0.20. The current screen and frozen randomized component pair remain unchanged. This grid includes points at and around efficacy and toxicity boundaries, spanning the three official truth regions. Only the S8.1 anchors impose formal calibration constraints; additional evaluated points do not alter selection or inherit anchor-level guarantees. Full-region results are shown in Supplementary Table S5 and Supplementary Figure S2A.

## S12.3 Frontline full-region grid

The frontline grid crosses true risk differences $\{0, 3/40, 3/20, 5/22\}$ with frontline experimental toxicity truths $\{0.10, 0.20, 0.30, 0.35\}$, giving 16 scenarios. The middle effects are equivalently 0.075 and 0.15. At R/R control truth 0.20, frontline control truth 0.50, $\lambda = 1$, and $\eta_T = 0$, corresponding R/R experimental efficacy truths are, in order, 0.20, $23/91$, $13/41$, and 0.40. Full-region results are shown in Supplementary Table S6 and Supplementary Figure S2B.

These current efficacy truths are obtained by inverse mapping of the prescribed frontline effects through the reference bridge. Rational values are retained at full precision rather than rounded before current outcome generation. With the reference safety bridge, current and frontline toxicity truths coincide. The broader grid is descriptive outside S8.2 anchors; exact boundary points retain their clinical classification.

## S12.4 De-escalation grid

The standalone grid evaluates losses $\{0, 0.02, 0.05, 0.10\}$, with standard efficacy truth 0.95 and satisfied burden gate. Each scenario retains the same current sample size, future noninferiority design, and frozen scalar threshold. This finite grid characterizes behavior at evaluated losses; it is not used to estimate a continuous transition point or universal clinical cutoff. See Supplementary Table S7 and Supplementary Figure S1.

# S13. Monte Carlo settings and numerical computation

Formal evaluations use 2,000 outer replicates per scenario and 1,000 inner predictive simulations per action when calculation is required. Higher-precision targeted verification uses 10,000 outer and 5,000 inner replicates. Outer repetitions quantify current-data variability and pathway behavior; inner repetitions approximate predictions conditional on each observed current dataset. These two sources of simulation variability are not interchangeable.

For a reported operating-characteristic proportion,

$$MCSE(\hat{p}) = \sqrt{\frac{\hat{p}(1-\hat{p})}{n}}.$$

The denominator is the applicable outer replicate count for unconditional proportions, including Fork passage, marginal support, joint patterns, and unconditional decision errors. For Fork-conditional support or randomized false discard it is the observed number of Fork-reaching trials at that scenario. A zero conditional denominator is not estimable, not a zero error probability. For continuous summaries such as expected current sample size or decision time, Monte Carlo SE is the replicate standard deviation divided by the square root of the outer replicate count.

All scenarios co-attaining a worst-case error are retained with their scenario-specific MCSEs. There is no separately defined MCSE for the maximum operator. Empirical zero events are reported as empirical zeros: the plug-in MCSE formula can also yield zero, but neither establishes a mathematically zero underlying probability.

## S13.1 Posterior risk-difference probabilities

For independent posterior probabilities $X \sim \text{Beta}(a_X, b_X)$ and $Y \sim \text{Beta}(a_Y, b_Y)$, a risk-difference probability is evaluated by one-dimensional integration:

$$P(X - Y > d) = \int_0^1 f_X(x) F_Y(x - d)\, dx.$$

Here $f_X$ is the Beta density and $F_Y$ the Beta cumulative distribution function, extended as zero below zero and one above one. The same formula covers positive superiority margins and the negative noninferiority margin. For a positive margin it is equivalently integrated from that margin to one. Continuous posterior distributions make equality at the risk-difference boundary probability zero.

Deterministic adaptive integration uses absolute and relative tolerances $1 \times 10^{-10}$. No extra Monte Carlo sampling is used to evaluate a future posterior risk difference. Count-indexed tables precompute posterior probabilities and success indicators for each possible future response-count pair. Randomized efficacy tables span counts 0 through 30 in each arm; de-escalation tables span 0 through 60. Experimental safety uses a separate toxicity-count table. LSA combined-data success is similarly indexed by added efficacy and toxicity counts, conditional on observed current counts.

Tables are reused for identical design, prior, margin, and success-threshold inputs. This separates deterministic formal-analysis success from the Monte Carlo future-data generation and avoids numerical variation from repeating posterior calculations inside every replicate.

# S14. Random-number generation and reproducibility

Simulation uses the L'Ecuyer-CMRG generator (L'Ecuyer, 1999) with a fixed configurable master seed and deterministic independent streams/substreams. Allocation distinguishes scenario, outer replicate, action, and sensitivity configuration. Distinct action streams preserve separate predictive calculations while the integrated workflow shares the same observed current dataset across its actions.

Stream assignments are fixed independently of serial or parallel scheduling, so changing execution order or worker assignment does not change a replicate's scientific draws. Reproducibility retains the scientific configuration, scenario definitions, stream-identifying seed metadata, and recorded R/package environment. No software version is inferred from an unrecorded environment.

Independent targeted-verification runs use a reserved, disjoint stream allocation rather than recycling the corresponding formal simulation streams. Common random numbers are not reused across sensitivity configurations. Deterministic numerical lookup construction does not consume simulation random numbers, and optional joint summaries are derived from existing paired indicators rather than new simulations. These controls preserve separation between current observation generation, action-specific prediction, and independent verification.

# S15. Interpretation boundaries

Predictive support is conditional on the specified future design, analysis, benchmarks, and bridges. Transport assumptions are assumed rather than empirically validated. Formal error-control claims apply to the finite calibration sets; expanded full-region grids are descriptive. Empirical zeros do not establish structural zeros, and absence of a support pattern in finite simulation does not impose an ordering among actions. E2D provides nonexclusive action-specific evidentiary support without a utility function or clinical-action ranking; choosing among supportable actions remains multidisciplinary judgment.

# Technical Supplement References

# Supplementary Tables S1–S8

## Table S1. Illustrative current-trial screening and safety rules

| Component | Model / criterion | Interim rule | Final rule | Calibration target / interpretation |
|---|---|---|---|---|
| Efficacy | Binary favorable response; binomial likelihood; Beta(1,1) posterior for prediction; Simon optimal two-stage design | $n_1 = 12,\ r_1 = 2$; continue for efficacy if >2 responses among 12 | $N = 25,\ r = 7$; efficacy passage requires $\geq 8$ responses among 25 | $p_0 = 0.20,\ p_1 = 0.40$; one-sided type I error $\leq 0.10$, power $\geq 0.80$; minimize null expected sample size; no additional Bayesian current efficacy gate |
| Safety | Binary unacceptable toxicity; binomial likelihood; Beta(1,1) prior; $T_{\max} = 0.30$ | Stop if $P(p_T > 0.30 \mid D_1) \geq 0.655$; operationally $\geq 5$ toxicities among 12 | Pass if $P(p_T < 0.30 \mid D) \geq 0.540$; operationally $\leq 6$ toxicities among 25 | Exact standalone safety-only sequential calibration: acceptable toxicity anchor 0.20, unacceptable anchor 0.40; interim stopping at 0.20 $\leq 0.10$; overall sequential safety passage at 0.40 $\leq 0.10$ |
| Development Fork | Requires interim continuation and both final gates | Continue only if efficacy and safety permit continuation; complete Stage-1 assessment before review | Both final efficacy and final safety must pass after complete assessment | All three requirements are necessary; non-Fork trials receive no support for the three post-RP2D actions |

*Notes.* Efficacy and toxicity failures are retained separately when overlapping; safety precedence applies only to mutually exclusive failure labels. Accrual pauses during Stage-1 assessment. De-escalation is standalone and does not use the Development Fork. Current Beta posteriors are used for prediction; they do not replace the Simon efficacy screen.

## Table S2. Action-specific future-study definitions and evidentiary-support rules

| Action | Future study | Predictive bridge / benchmark | Future efficacy success | Future safety / burden success | Predictive quantity used for support | Frozen current-evidence support threshold | Development Fork required? |
|---|---|---|---|---|---|---|---|
| Larger single-arm expansion (LSA) | Add 20 patients in the same population, setting, dose, and endpoint context; combine current + future data | Identity bridge for efficacy and toxicity; no between-cohort drift | Pr(pE > 0.20 \| Dcombined) ≥ 0.90 | Pr(pT < 0.30 \| Dcombined) ≥ 0.90 | Scalar PP(LSA) of complete efficacy AND safety success | PP(LSA) ≥ 0.20 | Yes |
| Same-setting randomized Phase II | Standalone N = 60, 30/arm; current patients excluded from future formal analysis | Experimental efficacy/toxicity identity bridges; independent control predictive benchmark Beta(2,8), mean 0.20, ESS 10 | Pr(pE − pC > 0.15 \| Dfuture) ≥ 0.90 | Pr(pT,E < 0.30 \| Dfuture,T,E) ≥ 0.90 | Paired PP(E) and PP(T); AND gate | PP(E) ≥ 0.16 AND PP(T) ≥ 0.41 | Yes |
| Frontline randomized Phase II | Standalone N = 60, 30/arm; current patients excluded from future formal analysis | Independent R/R benchmark Beta(2,8) and frontline benchmark Beta(5,5); log-odds efficacy transport with $\lambda = 1$; toxicity log-odds shift with $\eta_T = 0$ | Pr(pEF − pCF > 0.15 \| Dfuture) ≥ 0.90 | Pr(pT,F < 0.30 \| Dfuture,T,F) ≥ 0.90 | Paired PP(E) and PP(T); AND gate | PP(E) ≥ 0.27 AND PP(T) ≥ 0.39 | Yes |
| De-escalation | Current standard evidence n = 40; standalone future NI trial N = 120, 60/arm | Current standard Beta posterior; pDE = pSTD − d, truncated to [0,1] before generation; prespecified loss d | Pr(pDE − pSTD > −0.05 \| Dfuture) ≥ 0.90 | Burden-reduction criterion fixed as satisfied; not a simulated endpoint | Scalar PP(DE) of complete NI AND burden success | PP(DE) ≥ 0.17 | No |

*Notes.* Future-study posterior success thresholds (0.90) are distinct from current-evidence predictive-support thresholds. Standalone randomized future analyses use neutral independent Beta(1,1) priors without borrowing current observations or predictive benchmark priors. LSA uses the original Beta(1,1) priors updated once with combined data. Randomized safety concerns only the experimental arm; control-arm toxicity is neither simulated nor used. Joint future-success PP may be reported diagnostically but is not the randomized support gate. R/R denotes relapsed/refractory; ESS, effective sample size; NI, noninferiority.

## Table S3. Formal calibration anchors and error tolerances

| Action | Insufficient calibration anchors | Worthwhile calibration anchors | Maximum FA | Maximum FD | FD denominator | Notes |
|---|---|---|---|---|---|---|
| LSA | (0.15, 0.10); (0.15, 0.35); (0.25, 0.35); (0.40, 0.35); (0.55, 0.35) | (0.40, 0.10); (0.55, 0.10) | 0.20 | 0.20 | All current-trial replicates; end-to-end | Pairs are (pE, pT); Fork failure contributes to FD |
| Same-setting randomized Phase II | (0.15, 0.10); (0.15, 0.35); (0.25, 0.35); (0.40, 0.35); (0.55, 0.35) | (0.40, 0.10); (0.55, 0.10) | 0.10 | 0.20 | Fork-reaching trials | Pairs are (pE, pT); fixed control truth pC = 0.20; FA is unconditional end-to-end |

| Action | Insufficient calibration anchors | Worthwhile calibration anchors | Maximum FA | Maximum FD | FD denominator | Notes |
|---|---|---|---|---|---|---|
| Frontline randomized Phase II | (0, 0.10); (0, 0.35); (5/22, 0.35) | (5/22, 0.10) | 0.05 | 0.20 | Fork-reaching trials | Pairs are $(\Delta_F, p_{T,F})$; control truths 0.20 (R/R) and 0.50 (frontline), $\lambda = 1,\ \eta_T = 0$; FA is unconditional end-to-end |
| De-escalation | d = 0.05; 0.10 | d = 0; 0.02 | 0.05 | 0.20 | All standalone replicates | No Fork; burden gate fixed satisfied; standard efficacy truth 0.95 |

*Notes.* FA denotes false advance; FD denotes false discard. Gray-zone scenarios are not formal calibration anchors. In the original efficacy–toxicity grid, (0.25, 0.10) is gray-zone for LSA and same-setting randomized Phase II. Full-region evaluations are descriptive outside the listed anchors; formal error-control claims do not extend to all insufficient or worthwhile truths. De-escalation anchors are the listed finite losses, not a continuously estimated cutoff.

# Table S4. Frozen calibration rules and independent targeted verification

| Action | Frozen threshold(s) | Point FA | MCSE(FA) | Point FD | MCSE(FD) | Robust FA | Robust FD | Verification conclusion |
|---|---|---|---|---|---|---|---|---|
| LSA | c = 0.20 | 0.1620 | 0.01165 | 0.1938 | 0.00395274 | NA | NA | PASS |
| Same-setting randomized Phase II | cE = 0.16; cT = 0.41 | 0.0370 | 0.00597 | 0.08322935 | 0.00309 | 0.04893834 | 0.09201009 | PASS |
| Frontline randomized Phase II | cE = 0.27; cT = 0.39 | 0.0421 | 0.00201 | 0.16466015 | 0.00415 | 0.04662325 | 0.17296664 | PASS |
| De-escalation | c = 0.17 | 0.0000 | 0.0000 | 0.0000 | 0.0000 | NA | NA | PASS |

*Notes.* Values summarize the frozen final calibration frontier, incorporating prescribed scenario-specific higher-precision replacements. Point FA and FD are worst-case point estimates over the applicable finite sets. Randomized FA is unconditional end-to-end; randomized FD is conditional on reaching the Development Fork. LSA FD is end-to-end and includes Fork failure; de-escalation is standalone. Robust randomized quantities take the maximum of anchor-specific estimate + 2×MCSE after adjustment at each anchor. Point and robust maxima can occur at different anchors, so robust columns are not necessarily point-column values plus twice the displayed MCSE. NA indicates that the randomized robust criterion does not apply to scalar actions. Empirical zero errors and their zero plug-in MCSEs do not establish mathematical zero probabilities. Independent targeted verification used 10,000 outer and 5,000 inner replicates for the six prespecified settings, assessed the frozen rules, and did not reopen threshold optimization. Failure of a frozen rule would trigger scientific adjudication, not automatic recalibration or replacement.

# Table S5. Full-region same-setting randomized Phase II operating characteristics

Control truth is fixed at pC = 0.20; frozen component thresholds are cE = 0.16 and cT = 0.41.

| $p_E$ | $p_T$ | Truth-region class | Fork probability | $P(PP_E \geq 0.16 \mid \text{Fork})$ | $P(PP_T \geq 0.41 \mid \text{Fork})$ | Overall support | $P(\text{support} \mid \text{Fork})$ | MCSE(overall support) | Calibration anchor? |
|---|---|---|---|---|---|---|---|---|---|
| 0.15 | 0.10 | Insufficient | 0.0275 | 0.982 | 0.873 | 0.0235 | 0.855 | 0.0034 | Insufficient |
| 0.15 | 0.20 | Insufficient | 0.0130 | 1.000 | 0.500 | 0.0065 | 0.500 | 0.0018 | No |
| 0.15 | 0.30 | Insufficient | 0.0035 | 1.000 | 0.429 | 0.0015 | 0.429 | 0.0009 | No |
| 0.15 | 0.35 | Insufficient | 0.0025 | 0.800 | 0.200 | 0.0005 | 0.200 | 0.0005 | Insufficient |
| 0.20 | 0.10 | Insufficient | 0.1010 | 0.975 | 0.926 | 0.0910 | 0.901 | 0.0064 | No |
| 0.20 | 0.20 | Insufficient | 0.0855 | 0.965 | 0.573 | 0.0475 | 0.556 | 0.0048 | No |
| 0.20 | 0.30 | Insufficient | 0.0250 | 1.000 | 0.180 | 0.0045 | 0.180 | 0.0015 | No |
| 0.20 | 0.35 | Insufficient | 0.0180 | 0.972 | 0.222 | 0.0040 | 0.222 | 0.0014 | No |
| 0.25 | 0.10 | Gray-zone | 0.2450 | 0.976 | 0.900 | 0.2145 | 0.876 | 0.0092 | No |
| 0.25 | 0.20 | Gray-zone | 0.2075 | 0.976 | 0.607 | 0.1225 | 0.590 | 0.0073 | No |
| 0.25 | 0.30 | Insufficient | 0.0805 | 0.975 | 0.248 | 0.0190 | 0.236 | 0.0031 | No |
| 0.25 | 0.35 | Insufficient | 0.0495 | 0.990 | 0.323 | 0.0160 | 0.323 | 0.0028 | Insufficient |
| 0.35 | 0.10 | Worthwhile | 0.6415 | 0.991 | 0.927 | 0.5890 | 0.918 | 0.0110 | No |
| 0.35 | 0.20 | Worthwhile | 0.5405 | 0.997 | 0.595 | 0.3210 | 0.594 | 0.0104 | No |
| 0.35 | 0.30 | Insufficient | 0.2110 | 0.988 | 0.287 | 0.0595 | 0.282 | 0.0053 | No |
| 0.35 | 0.35 | Insufficient | 0.1205 | 0.983 | 0.224 | 0.0270 | 0.224 | 0.0036 | No |
| 0.40 | 0.10 | Worthwhile | 0.8025 | 0.996 | 0.924 | 0.7385 | 0.920 | 0.0098 | Worthwhile |
| 0.40 | 0.20 | Worthwhile | 0.6350 | 0.996 | 0.570 | 0.3605 | 0.568 | 0.0107 | No |
| 0.40 | 0.30 | Insufficient | 0.2745 | 0.995 | 0.350 | 0.0955 | 0.348 | 0.0066 | No |
| 0.40 | 0.35 | Insufficient | 0.1450 | 1.000 | 0.300 | 0.0435 | 0.300 | 0.0046 | Insufficient |
| 0.55 | 0.10 | Worthwhile | 0.9755 | 1.000 | 0.925 | 0.9020 | 0.925 | 0.0066 | Worthwhile |
| 0.55 | 0.20 | Worthwhile | 0.7575 | 1.000 | 0.574 | 0.4350 | 0.574 | 0.0111 | No |
| 0.55 | 0.30 | Insufficient | 0.3330 | 1.000 | 0.321 | 0.1070 | 0.321 | 0.0069 | No |
| 0.55 | 0.35 | Insufficient | 0.1685 | 1.000 | 0.252 | 0.0425 | 0.252 | 0.0045 | Insufficient |

*Notes.* Each row contains 2,000 formal outer replicates, with 1,000 inner simulations when prediction is required. Component-gate probabilities and conditional overall support use Fork-reaching trials as their denominator; they are not future-study posterior success probabilities. Overall support is unconditional end-to-end and requires Fork passage plus both component gates. Four decimal places are retained for unconditional proportions to preserve 1/2,000 increments; conditional probabilities are shown to three decimals and support MCSEs to four. Gray-zone is the manuscript label for the frozen output's gray class. Formal error-control claims apply only to prespecified calibration anchors, not all 24 scenarios.

# Table S6. Full-region frontline randomized Phase II operating characteristics

Reference control truths are 0.20 (R/R) and 0.50 (frontline), with $\lambda = 1$ and $\eta_T = 0$. Frozen component thresholds are cE = 0.27 and cT = 0.39.

| $\Delta_F$ | Corresponding $p_{ER}$ | $p_{T,F}$ | Truth-region class | Fork probability | $P(PP_E \geq 0.27 \mid \text{Fork})$ | $P(PP_T \geq 0.39 \mid \text{Fork})$ | Overall support | $P(\text{support} \mid \text{Fork})$ | MCSE(overall support) | Calibration anchor? |
|---|---|---|---|---|---|---|---|---|---|---|
| 0 | 0.20 | 0.10 | Insufficient | 0.0850 | 0.447 | 0.935 | 0.0355 | 0.418 | 0.0041 | Insufficient |
| 0 | 0.20 | 0.20 | Insufficient | 0.0655 | 0.504 | 0.656 | 0.0205 | 0.313 | 0.0032 | No |
| 0 | 0.20 | 0.30 | Insufficient | 0.0250 | 0.480 | 0.500 | 0.0065 | 0.260 | 0.0018 | No |
| 0 | 0.20 | 0.35 | Insufficient | 0.0130 | 0.538 | 0.346 | 0.0035 | 0.269 | 0.0013 | Insufficient |
| 3/40 | 23/91 | 0.10 | Gray-zone | 0.2705 | 0.558 | 0.954 | 0.1430 | 0.529 | 0.0078 | No |
| 3/40 | 23/91 | 0.20 | Gray-zone | 0.2055 | 0.584 | 0.703 | 0.0870 | 0.423 | 0.0063 | No |
| 3/40 | 23/91 | 0.30 | Insufficient | 0.0860 | 0.512 | 0.471 | 0.0220 | 0.256 | 0.0033 | No |
| 3/40 | 23/91 | 0.35 | Insufficient | 0.0460 | 0.609 | 0.424 | 0.0110 | 0.239 | 0.0023 | No |
| 3/20 | 13/41 | 0.10 | Worthwhile | 0.5150 | 0.732 | 0.959 | 0.3605 | 0.700 | 0.0107 | No |
| 3/20 | 13/41 | 0.20 | Worthwhile | 0.4245 | 0.729 | 0.675 | 0.2095 | 0.494 | 0.0091 | No |
| 3/20 | 13/41 | 0.30 | Insufficient | 0.1835 | 0.730 | 0.477 | 0.0615 | 0.335 | 0.0054 | No |
| 3/20 | 13/41 | 0.35 | Insufficient | 0.0945 | 0.757 | 0.386 | 0.0295 | 0.312 | 0.0038 | No |
| 5/22 | 0.40 | 0.10 | Worthwhile | 0.8120 | 0.874 | 0.952 | 0.6740 | 0.830 | 0.0105 | Worthwhile |
| 5/22 | 0.40 | 0.20 | Worthwhile | 0.6340 | 0.883 | 0.695 | 0.3935 | 0.621 | 0.0109 | No |
| 5/22 | 0.40 | 0.30 | Insufficient | 0.2505 | 0.870 | 0.465 | 0.1035 | 0.413 | 0.0068 | No |
| 5/22 | 0.40 | 0.35 | Insufficient | 0.1365 | 0.853 | 0.392 | 0.0435 | 0.319 | 0.0046 | Insufficient |

*Notes.* Exact rational efficacy values are retained. Each row contains 2,000 formal outer replicates, with 1,000 inner simulations when prediction is required. Component-gate and conditional-support probabilities are conditional on reaching the Fork; overall support is unconditional end-to-end. Display precision follows Table S5. Results are conditional on the specified transport model with $\lambda = 1$ and $\eta_T = 0$; the simulation does not empirically validate transportability. Formal calibration constraints apply only to listed anchors; other full-region scenarios are descriptive, including non-anchor boundary points.

# Table S7. De-escalation operating characteristics

| Efficacy loss d | pSTD | pDE | Truth-region class | 5th percentile of $PP_{\text{DE}}$ | Median $PP_{\text{DE}}$ | 95th percentile of $PP_{\text{DE}}$ | Support probability | Number supported / 2,000 | Calibration anchor class |
|---|---|---|---|---|---|---|---|---|---|
| 0.00 | 0.95 | 0.95 | Worthwhile | 0.330 | 0.452 | 0.666 | 1.000 | 2000/2000 | Worthwhile |
| 0.02 | 0.95 | 0.93 | Worthwhile | 0.214 | 0.267 | 0.316 | 1.000 | 2000/2000 | Worthwhile |
| 0.05 | 0.95 | 0.90 | Insufficient | 0.084 | 0.101 | 0.118 | 0.000 | 0/2000 | Insufficient |
| 0.10 | 0.95 | 0.85 | Insufficient | 0.008 | 0.016 | 0.026 | 0.000 | 0/2000 | Insufficient |

*Notes.* This is a standalone demonstration, without Development Fork conditioning, using current standard evidence n = 40, future N = 120 with 60 per arm, and frozen support threshold cDE = 0.17. The burden criterion is fixed as satisfied. The empirical transition between evaluated losses 0.02 and 0.05 does not define a universal or continuously estimated clinical cutoff. Observed support of 1.000 or 0.000 describes finite simulations, not mathematical certainty.

# Table S8. Complete post-RP2D support-pattern probabilities

Support-bit order is LSA – same-setting RCT – frontline RCT. Probabilities are unconditional over the complete current-trial pathway.

| pE | pT | 000 | 001 | 010 | 011 | 100 | 101 | 110 | 111 |
|---|---|---|---|---|---|---|---|---|---|
| 0.15 | 0.10 | 0.9750 | 0.0000 | 0.0000 | 0.0000 | 0.0010 | 0.0005 | 0.0135 | 0.0100 |
| 0.15 | 0.35 | 0.9980 | 0.0000 | 0.0000 | 0.0000 | 0.0010 | 0.0000 | 0.0005 | 0.0005 |
| 0.25 | 0.10 | 0.7515 | 0.0000 | 0.0000 | 0.0000 | 0.0220 | 0.0050 | 0.0875 | 0.1340 |
| 0.25 | 0.35 | 0.9625 | 0.0000 | 0.0000 | 0.0000 | 0.0280 | 0.0030 | 0.0035 | 0.0030 |
| 0.40 | 0.10 | 0.1885 | 0.0000 | 0.0000 | 0.0000 | 0.0380 | 0.0240 | 0.0950 | 0.6545 |
| 0.40 | 0.35 | 0.8570 | 0.0000 | 0.0000 | 0.0000 | 0.0845 | 0.0235 | 0.0040 | 0.0310 |

| pE | pT | 000 | 001 | 010 | 011 | 100 | 101 | 110 | 111 |
|---|---|---|---|---|---|---|---|---|---|
| 0.55 | 0.10 | 0.0295 | 0.0000 | 0.0000 | 0.0000 | 0.0460 | 0.0245 | 0.0120 | 0.8880 |
| 0.55 | 0.35 | 0.8210 | 0.0000 | 0.0000 | 0.0000 | 0.1075 | 0.0285 | 0.0035 | 0.0395 |

*Notes.* All eight three-bit patterns are permitted by E2D. Non-Fork trials contribute 000, as do Fork-reaching trials with no supported action. Patterns 001, 010, and 011 were not observed in these finite simulations; their empirical absence does not represent a structural hierarchy or logical impossibility. Each scenario contains 2,000 formal outer replicates. Four decimal places preserve the frozen pattern probabilities and permit exact recovery of marginal support within arithmetic precision.

# Supplementary Figures S1–S2

**Figure provenance note.** The figures below are assembled directly from frozen supplementary-table quantities. Earlier production planning also contemplated candidate-threshold calibration-landscape graphics requiring retained candidate-frontier objects. Those raw frontier objects are not embedded in the present technical document, so no landscape is reconstructed from selected-rule summaries alone. Table S4 remains the authoritative report of frozen calibration and independent verification.

## Supplementary Figure S1. De-escalation predictive support at prespecified efficacy losses

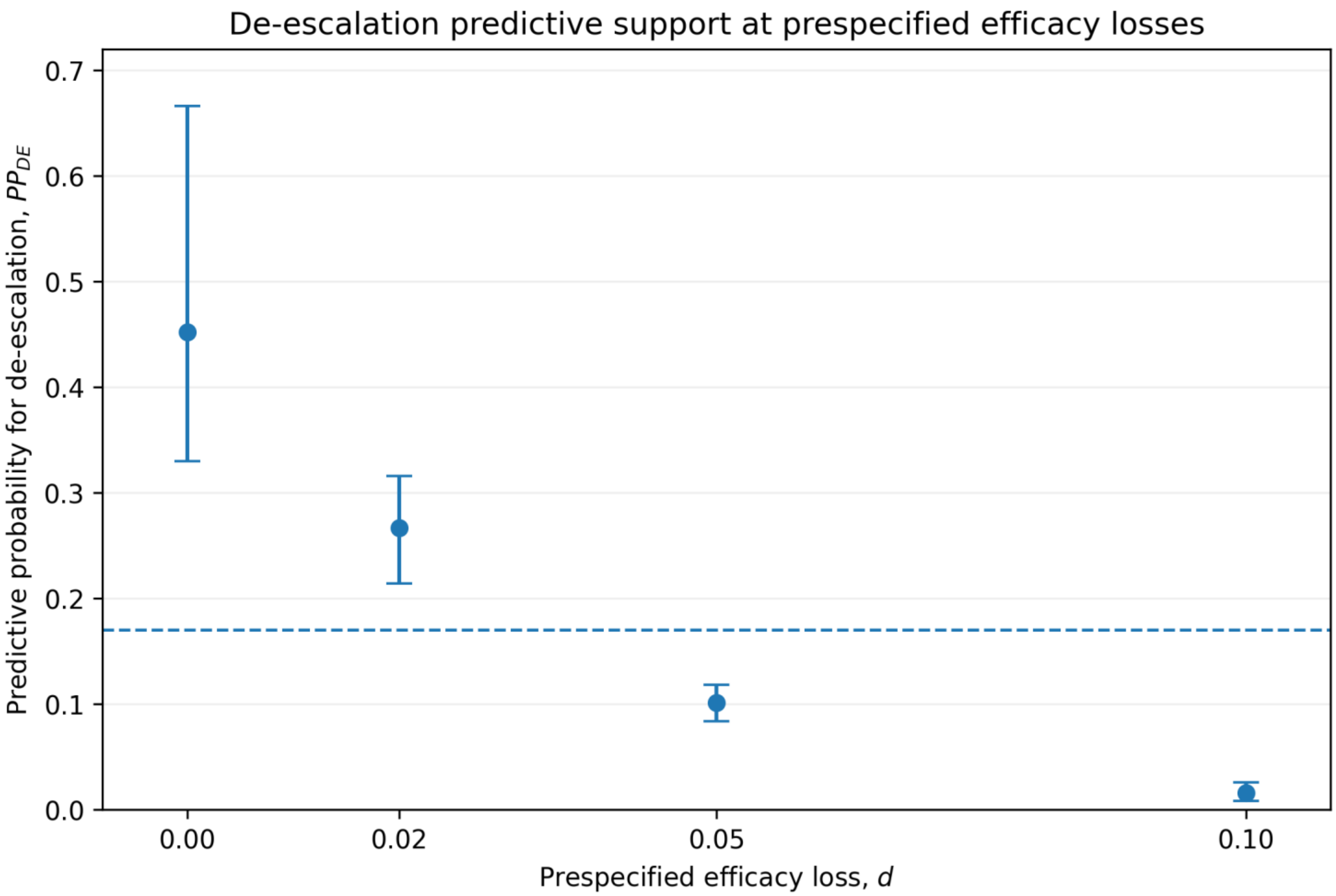

Points show the median predictive probability for the de-escalation action across current-data replicates; vertical intervals show the 5th–95th percentile range at each prespecified efficacy loss. The dashed horizontal line marks the frozen current-evidence support threshold cDE = 0.17, not the future noninferiority posterior criterion. The standalone de-escalation demonstration does not use the Development Fork. Values are copied from Supplementary Table S7. The apparent transition between evaluated losses does not define a universal or continuously estimated clinical cutoff.

## Supplementary Figure S2. Broader randomized support across evaluated efficacy–toxicity grids

### Panel A. Same-setting randomized Phase II

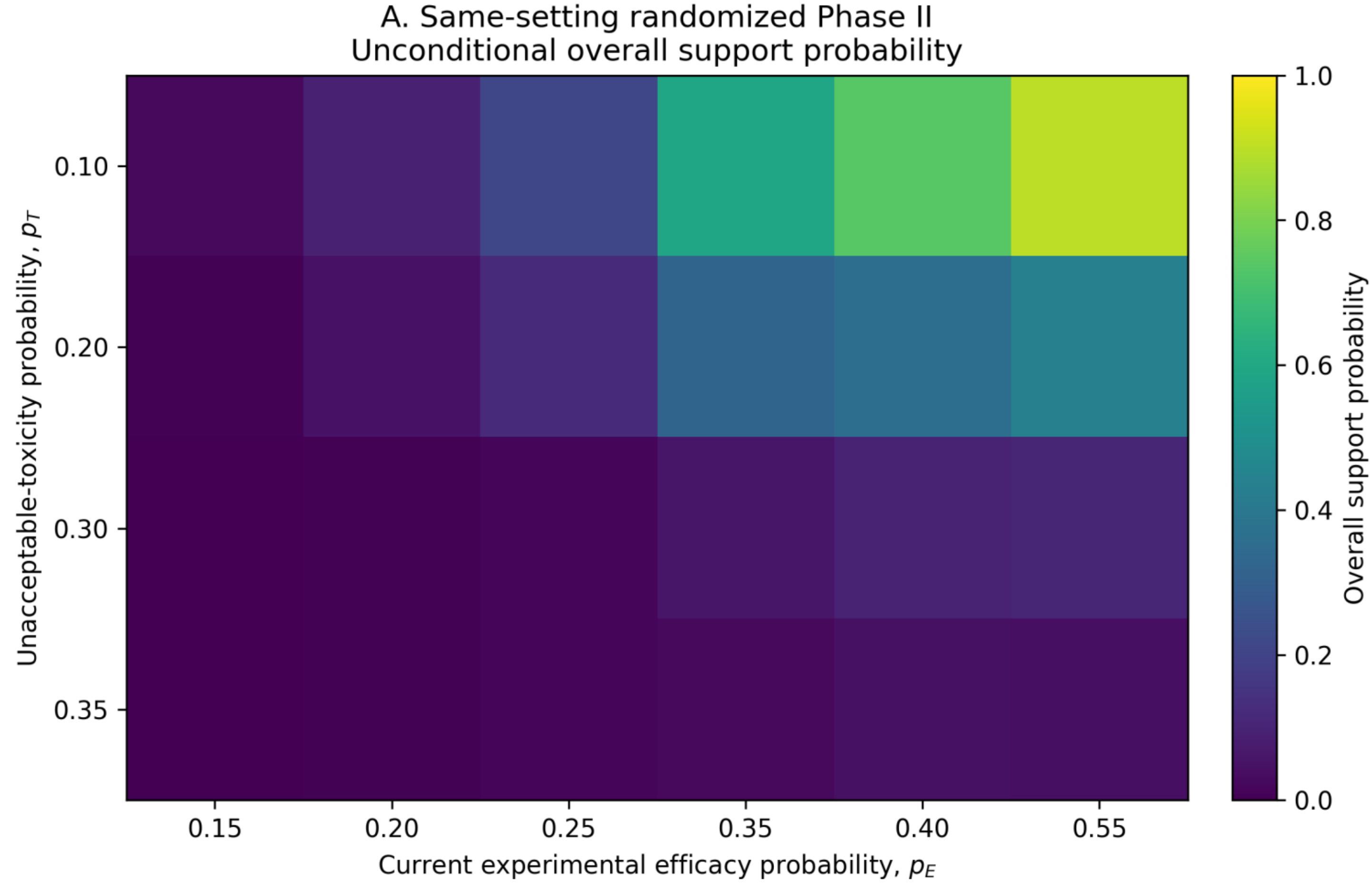

## Panel B. Frontline randomized Phase II

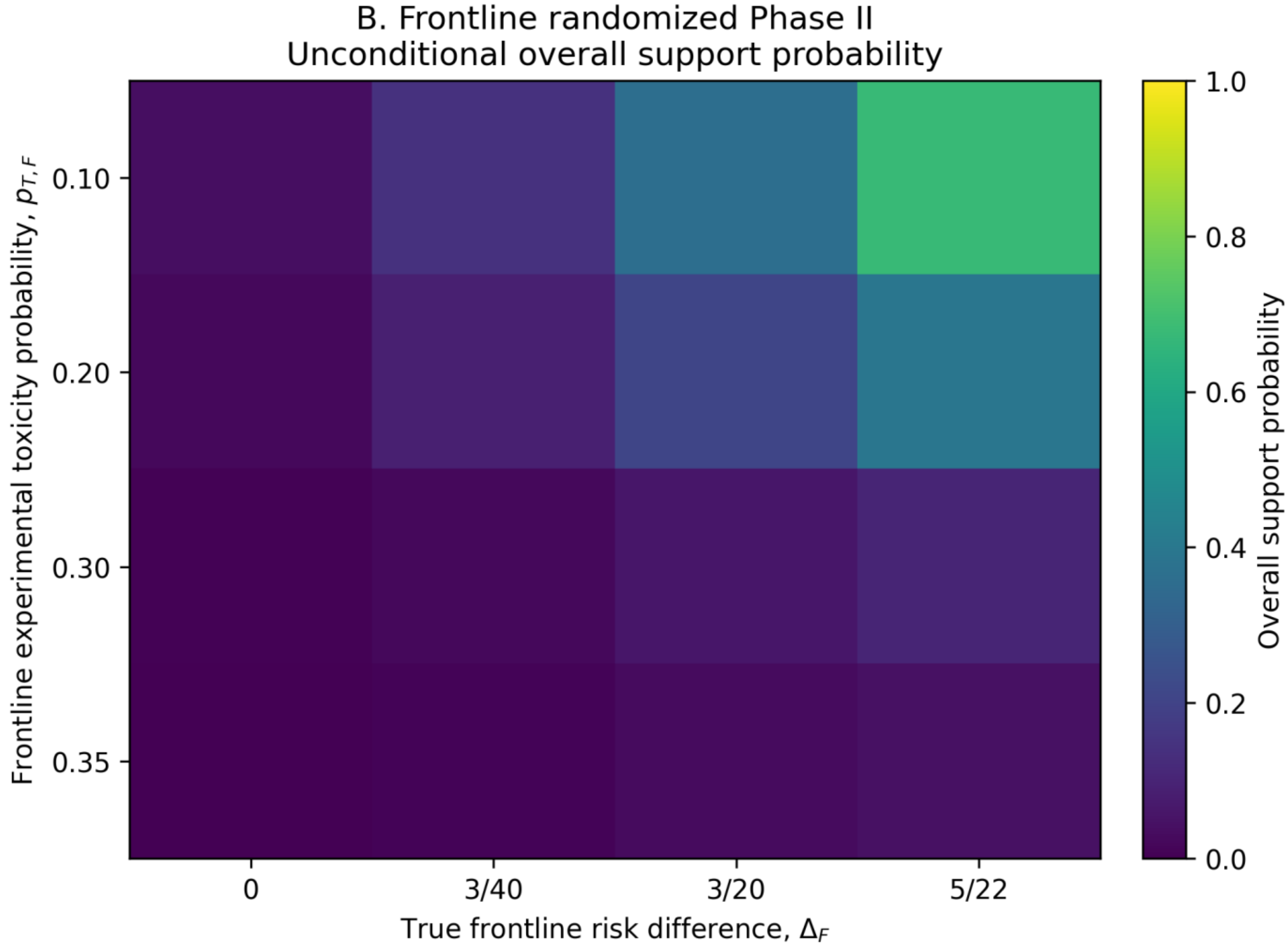

Heatmaps show unconditional end-to-end evidentiary-support probabilities for two contemplated randomized actions. Panel A shows same-setting randomized Phase II across the expanded current efficacy–toxicity grid; control truth is fixed at 0.20 and the frozen component thresholds are cE = 0.16 and cT = 0.41. Panel B shows frontline randomized Phase II across true frontline risk differences and frontline experimental toxicity under the reference transport model with λ = 1 and ηT = 0; frozen component thresholds are cE = 0.27 and cT = 0.39. Values correspond to Supplementary Tables S5 and S6. These full-region results are descriptive outside prespecified calibration anchors; formal error-control claims do not extend to the continuous parameter space, and the frontline panel does not empirically validate transportability.

**Assembly boundary.** The revised main Figure 2 and revised main Table 3 are main-manuscript presentation assets and are therefore not duplicated here. Their joint support-pattern quantities are fully recoverable from Supplementary Table S8.